\documentclass[aps,prd,twocolumn,nofootinbib,floatfix]{revtex4}
\usepackage{subfigure}
\usepackage{graphicx}
\usepackage{dcolumn}
\usepackage{epsfig}
\usepackage{epstopdf}
\usepackage{amsmath}
\usepackage{amsfonts}
\usepackage{amssymb}
\usepackage{mathtools}
\begin{document}

\title{Constraints on dark matter scenarios from measurements of the galaxy luminosity function at high redshifts}
\author{P.S. Corasaniti$^{1}$, S. Agarwal$^{3,2,1}$, D.J.E. Marsh$^{4}$, S. Das$^{5}$}
\affiliation{$^{1}$LUTH, UMR 8102 CNRS, Observatoire de Paris, PSL Research University, Universit\'e Paris Diderot, 92190 Meudon, France \\
$^{2}$African Institute for Mathematical Sciences, 6--8 Melrose Rd, Muizenberg
7945, South Africa \\
$^{3}$Stellenbosch University, South Africa \\
$^{4}$King's College London, Strand, London, WC2R 2LS, United Kingdom \\
$^{5}$Indian Institute of Astrophysics, 560034 Bangalore, India}

\begin{abstract}
We use state-of-art measurements of the galaxy luminosity function (LF) at $z=6,7$ and $8$ to derive constraints on warm dark matter (WDM), late-forming dark matter (LFDM) and ultra-light axion dark matter (ULADM) models alternative to the cold dark matter (CDM) paradigm. To this purpose we have run a suite of high-resolution N-body simulations to accurately characterise the low-mass end of the halo mass function and derive DM model predictions of the high-z luminosity function. In order to  convert halo masses into UV-magnitudes we introduce an empirical approach based on halo abundance matching which allows us to model the LF in terms of the amplitude and scatter of the ensemble average star formation rate halo mass relation, $\langle {\rm SFR}({\rm M_{ h}},z)\rangle$, of each DM model. We find that independent of the DM scenario the average SFR at fixed halo mass increases from $z=6$ to $8$, while the scatter remains constant. At halo mass ${\rm M_{h}}\gtrsim 10^{12}\,{\rm M}_\odot$ h$^{-1}$ the average SFR as function of halo mass follows a double power law trend that is common to all models, while differences occur at smaller masses. In particular, we find that models with a suppressed low-mass halo abundance exhibit higher SFR compared to the CDM results. Thus, different DM models predict a different faint-end slope of the LF which causes the goodness-of-fit to vary within each DM scenario for different model parameters. Using deviance statistics we obtain a lower limit on the WDM thermal relic particle mass, $m_{\rm WDM}\gtrsim 1.5$ keV at $2\sigma$. In the case of LFDM models, the phase transition redshift parameter is bounded to $z_t\gtrsim 8\cdot 10^5$ at $2\sigma$. We find ULADM best-fit models with axion mass $m_a\gtrsim 1.6\cdot 10^{-22}$ eV to be well within $2\sigma$ of the deviance statistics. We remark that measurements at $z=6$ slightly favour a flattening of the LF at faint UV-magnitudes. This tends to prefer some of the non-CDM models in our simulation suite, although not at a statistically significant level to distinguish them from CDM. 
\end{abstract} 

\maketitle

\section{Introduction}\label{intro}
In the past few years there has been significant progress in the characterization of the high-redshift UV-luminosity function (LF) (see e.g.~\cite{Finkelstein2012,Bouwens2012,Oesch2013,Schenker2013,Mclure2013,Bouwens2015,Finkelstein2015,McLeod2015,Bouwens2016a}). Measurements from galaxy samples at $z\gtrsim 4$ have shown that the slope of LF remains steep to ${\rm M_{UV}}\sim -17$ magnitudes (see e.g. \cite{Bouwens2015,Finkelstein2015}) with important implications for scenarios of cosmic reionization. Evidence of such steepness persisting to very faint magnitudes (${\rm M_{UV}}\sim -13$) would imply the existence of a large population of dim galaxies contributing to the reionization of the universe (see e.g. \cite{Munoz2011,Jaacks2012,Cai2014,Mason2015}). However, it is only very recently that observations have begun probing the galaxy LF at such low UV-luminosities. As an example, measurements of the LF to ${\rm M_{UV}}\approx -15$ at $z\sim 6$ and ${\rm M_{UV}}\approx -17$ at $z\sim 8$ have been obtained in \cite{Ishigaki2015,Atek2015a,Atek2015b}, while estimates to even fainter magnitudes have been obtained by Livermore, Finkelstein and Lotz \cite{Livermore2016}. The latter have been able to characterise for the first time the LF to ${\rm M_{UV}}= -12.5$ at $z \sim 6$, ${\rm M_{UV}}=-14$ at $z \sim 7$ and ${\rm M_{UV}}=-15$ at $z \sim 8$, showing that the LF slope remains steep to very faint magnitudes and at high-redshifts. 

These measurements have been possible thanks to the detection of very faint high-redshift objects through the gravitational lensing magnification caused by massive galaxy clusters that are the targets of the {\it Hubble} Frontier Fields (HFF) program \cite{Coe2015,Lotz2016}. This novel approach is a promising alternative to deep galaxy survey searches such as the {\it Hubble} Ultra Deep Field \cite{Beckwith2006,Ellis2013,Illingworth2013}, but it is not exempt of systematic errors that can bias the determination of the LF. As an example, the uncertainty in the assumed size distribution of very faint galaxies \cite{Bouwens2016b} and the magnification error due to lens model uncertainties can alter the LF faint-end slope \cite{Bouwens2016c}. The latter has been shown to be the dominant source of systematics. In particular the analysis of \cite{Bouwens2016c} has indicated that when carefully assessed, current LF estimates cannot exclude the presence of a flattening of the LF at the faint-end, as expected from a number of numerical simulations studies \cite{Jaacks2013,Oshea2015}, which would call into question some of the proposed reionization scenarios. However, the implications of these measurements are far wider than probing the link between galaxy formation models and cosmic reionization history, since they provide a test of the nature of dark matter (DM) itself. 

In the standard cosmological model (e.g. \cite{Peebles1993}), DM consists of cold, collisionless particles interacting with visible matter (and indeed other DM particles and neutrinos) only via gravity. This is known as the cold dark matter (CDM) paradigm, and has various motivations from particle physics such as supersymmetry, extra dimensions, axions, and string theory (for reviews, see e.g. Refs.~\cite{1996PhR...267..195J,2005PhR...405..279B,2016PhR...643....1M}). In such a scenario the faint galaxies observed at high-redshift populate small-mass DM halos. As an example, analytical models of the LF suggest that galaxies with magnitude ${\rm M_{UV}}\approx -15$ at $z \sim 8$ should be hosted in halos of mass of $\approx 10^9\,M_\odot$ h$^{-1}$ (see e.g. \cite{Munoz2011,Cai2014}). Therefore, the recent measurements of the faint-end of the galaxy LF function at $z=6,7$ and $8$ probe the lightest and earliest to form DM objects, which are at the frontier of our knowledge.

The CDM paradigm has been tremendously successful at reproducing observations of the large-scale distribution of matter in the universe \cite{Spergel2003,Tegmark2004,Massey2007,Planck2015}. In contrast, the emergence of anomalies at small scales and the lack of detection of supersymmetric weakly interactive particles (WIMPs) or QCD axions in the lab (e.g. Refs.~\cite{2010PhRvL.104d1301A,pdg}) have prompted the investigation of broader scenarios that evade detection in standard channels. For example, direct detection interpretations are altered when the DM production method breaks the link between thermal cross-section and abundance, when production is non-thermal, or when symmetry dictates particular couplings to be absent or suppressed. In the present work we will be particularly interested in models of DM that not only evade direct detection, but also differ from CDM in terms of cosmological structure formation, and as such can be probed with the high-$z$ LF. In these scenarios, astrophysics offers a probe of DM particle physics complementary to laboratory based searches.

We will consider three examples of DM models fitting this prescription. A warm dark matter (WDM) component with thermal relic particle mass $m_{\rm WDM}$ $\mathcal{O}(\sim{\rm keV})$, inspired by particle physics models of sterile neutrinos, has been advocated as a solution to the small-scale anomalies of CDM (see e.g. \cite{WDMth,WDMph}). Sterile neutrinos in this mass range cannot be detected in standard WIMP searches at least with current experimental capabilities (see e.g. \cite{Liao2014,Mertens2015,Campos2016}), but leave imprints on the cosmic structure formation due to their thermal velocities. Ultralight axions (ULAs) with mass $\mathcal{O}(10^{-22}\text{ eV})$ may be present in hidden sectors, evading DM searches based on the couplings of the QCD axion~\cite{2006JHEP...06..051S,2010PhRvD..81l3530A,2016PhRvD..93b5027K}. ULAs and other models of scalar field/wave DM affect structure formation due to their large de Broglie wavelength~\cite{khlopov_scalar,1990PhRvL..64.1084P,Sin1994,Lee1996,2000PhRvL..85.1158H,2010PhRvD..82j3528M,2014MNRAS.437.2652M,Schive2014,Hui2016}. Our third and final benchmark model is late-forming dark matter (LFDM) \cite{Das2011,Agarwal2015}. In such a scenario DM particles emerge from a scalar field undergoing a phase transition near matter-radiation equality which alters the small scale distribution of density fluctuations.\footnote{Other related scenarios that we do not consider include self-interacting DM (e.g. Refs.~\cite{2000PhRvL..84.3760S,2014PhRvD..89c5009K}), the ``effective theory of structure formation''€™~\cite{2016PhRvD..93l3527C}, and generalised models~\cite{1998ApJ...506..485H,2016PhRvD..94b3510K,2016PhRvD..94d3512K}.}

A common feature of these scenarios is the suppression of matter density fluctuations below a cut-off scale that depends on the specificity of the DM particle model. Traces of this signature have been tightly constrained using matter power spectrum measurements at $z\sim 4-5$ from Lyman-$\alpha$ forest observations (see e.g. \cite{Viel2013,Baur2016}). However, it has been pointed out that such bounds may relax if the thermal evolution of the intergalactic medium is a non-monotonic function of redshift \cite{Garzilli2015}. A detailed discussion of other caveats pertaining the properties of the intergalactic medium that enter such analyses can be found in \cite{Hui2016}. Alternatively, DM models predicting a cut-off in the matter power spectrum can be constrained using measurements of the abundance of faint galaxies at high redshifts. This is because the suppression of power at small scale leads to suppressed abundance of low-mass halos. Similarly, constraints on DM scenarios can be inferred from measurements of the dark matter distribution in the local universe. Indeed, it was the discovery of small scale anomalies in the distribution of structures surrounding the Milky Way, such as the core-vs-cusp problem \cite{Moore1994,Kuzio2011}, the missing satellites problem \cite{Klypin1999,Moore1999} and the too-big-to-fail problem \cite{Boylan2011,Boylan2012} that have prompted the study of non-standard DM models such as WDM. Nevertheless, in the low redshift universe it is hard to disentangle whether such anomalies are the result of the non-standard properties of DM or the consequence of baryon feedback (see e.g. \cite{Libeskind2007,Maccio2010,Pontzen2012}). This is because the amplitude and nature of the baryonic processes that contribute to the shaping distribution of matter at small scales and at late times remains largely uncertain (see e.g. \cite{Panarubia2012,Garrison2013}). In the high-redshift universe on the other hand, baryonic processes are expected to be less complex, thus it is possible that measurements of the abundance of faint high-z galaxies hosted in low mass DM halos may provide more pristine insights on DM.

Constraints on the WDM models using earlier high-redshift LF measurements have already been obtained in numerous works in the literature. As an example, the authors of \cite{Pacucci2013} have derived constraints on WDM thermal relic mass from estimates of the high-redshift galaxy number density and found $m_{\rm WDM}\ge 0.9$ keV at $2\sigma$. Using LF measurements at $z\sim 8-10$ in combination with bounds on the optical depth parameter from \textit{Planck} the authors of \cite{Lapi2015} found that $m_{\rm WDM}\sim 2-3$ keV. Strong exclusion bounds with $m_{\rm WDM}\ge 2.4$ keV at $2\sigma$ have been recently obtained in \cite{Menci2016} using the LF faint-end data from \cite{Livermore2016}. Differently from these analyses, the authors of \cite{Schultz2014} have obtained constraints on WDM models using high-redshift measurements of the cumulative luminosity function resulting in $m_{\rm WDM}>1.3$ keV at $\sim 2\sigma$. 

A key assumption in the analysis of WDM models using LF measurements is the derivation of the relation between halo mass and UV-magnitude that is necessary to convert the N-body calibrated halo mass function into the LF model prediction. In \cite{Schultz2014} the authors have estimated this relation for the CDM model using halo abundance matching (HAM) and linearly extrapolated the relation to faint galaxy magnitudes. However, it is far from obvious that such a relation can be assumed to hold independent of the underlying DM model assumptions. For example, in \cite{Bozek2015} the authors have used a similar methodology to derive constraints on mixed axion-CDM models. However, unlike \cite{Schultz2014}, they calibrated halo mass and UV-magnitude relation for each of the investigated axion-CDM models. Instead of considering the cumulative LF, the authors of \cite{Schive2016} have used a conditional LF method to constrain wave dark matter models directly against LF measurements. Their analysis has found a lower bound on a boson-like DM particle mass, $m_{\psi}\ge 1.2\cdot 10^{-22}$ eV at $2\sigma$.  

\begin{figure}
\includegraphics[scale=0.43]{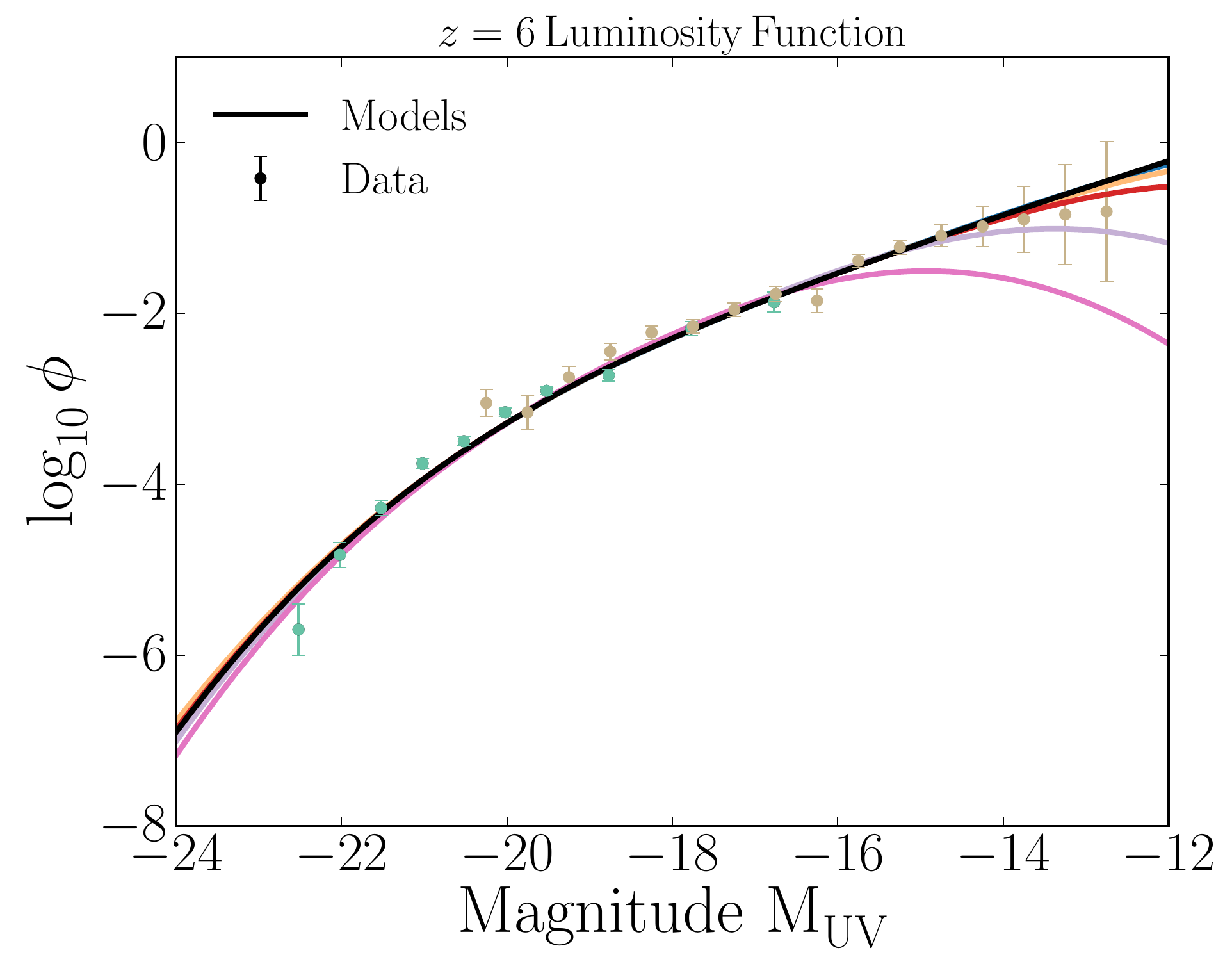}
\caption{WDM best-fit models against LF data at $z=6$. Curves from bottom to top corresponds to WDM-1 to WDM-5 respectively.}\label{figplotpres}
\end{figure}

Here, we aim to derive up-to-date bounds on several DM scenarios consisting of WDM, LFDM and ULADM, using a large compilation of high-redshift LF data. To this purpose we have run high-resolution N-body simulations which take as input modified initial conditions appropriate to each of the DM models considered. We accurately estimate the corresponding halo mass functions at low masses and at high-redshifts that we use to infer DM model predictions of the high-redshift luminosity function. As already stressed, this requires assuming a relation between halo mass and galaxy UV-luminosity. In order to improve upon the approaches of previous studies, we have developed a hybrid methodology (described in Section~\ref{method}) which intends to make progress in the use of LF measurements by addressing two important aspects previously overlooked: (i) it accounts for dust extinction on rest-frame UV photons which may alter the HAM inferred relation between halo mass and UV-magnitude at different redshifts. This is done by correcting the UV-luminosities using the established correlation between dust extinction and the UV-continuum slope \cite{Meurer1999}, and (ii) it allows us to gain insights on the DM model-dependence of the star formation rate (SFR) of high-redshift galaxies. This involves using the Kennicutt-relation \cite{Kennicutt1998} to convert the corrected UV LF measurements into SFR density functions (see e.g. \cite{Smit2012,Mason2015,Mashian2016}). 

Our analysis of the high-redshift galaxy LF indicates that in the case of WDM models the thermal relic mass is constrained to be $m_{\rm WDM}\gtrsim 1.5$ keV at $2\sigma$ (see Fig.~\ref{figplotpres} for a preview of the results at $z=6$). The same dataset excludes LFDM models with phase transition redshift $z_t\le 5\cdot 10^5$ at more than $4\sigma$, while ULADM models with $m_{\rm a}\gtrsim 1.6 \cdot 10^{-22}$ eV are compatible with the data within $2\sigma$. 

The paper is organised as follows: Section~\ref{nbody} describes the DM models, the N-body simulation characteristics and halo detection scheme. Section~\ref{method} details the hybrid method we use to model the high-redshift galaxy UV LF. The LF datasets used in this work are described in Section~\ref{data_analysis} with the results presented in Section~\ref{result}. We discuss and conclude in Section~\ref{conclude}.

\section{N-body simulations}\label{nbody}
In this section we describe the properties of the simulated DM models, the characteristics of the N-body simulations, the identification of halos and the evaluation of the halo mass functions.

\subsection{Cosmological models}
Our reference cosmological model is a standard flat cold dark matter model with cosmological constant (to which we will simply refer as CDM) specified by the following set of model parameters: matter density $\Omega_m=0.3$, baryon density $\Omega_b=0.046$, reduced Hubble parameter $h=0.7$, scalar spectral index $n_s=0.99$,\footnote{After running the simulations of the CDM, WDM and LFDM models we realised that we had inadvertently generated the linear power spectrum of our reference CDM model with the scalar spectral index set to $n_s=0.99$ rather than the \textit{Planck} best-fit value $n_s=0.96$. Given the limited computing time allocation available to us we were unable to re-run these models, therefore we decided to complete the simulation suite with $n_s=0.99$ for the ULADM models as well. By using a larger value of $n_s$ our numerical simulations systematically predict slightly more power at small scales, resulting in a slight increase in the abundance of small-mass halos. This has the tendency to relax the constraints on the alternative DM models. Hence, we obtain more conservative bounds on the DM model parameters than what we would have inferred using $n_s=0.96$.} and root-mean-square fluctuation amplitude at $8\,h^{-1} {\rm Mpc}$ $\sigma_8=0.8$. For this model we compute the linear matter power spectrum using the code {\sc camb} \cite{Lewis2000}. The cosmological parameters listed above are common to all simulated models. We consider three classes of DM: 
\begin{itemize}
\item WDM models consisting of realizations with thermal relic particle mass $m_\textrm{WDM}=0.696$ keV (WDM-1), $1.000$ keV (WDM-2), $1.465$ keV (WDM-3), $2.000$ keV (WDM-4) and $2.441$ keV (WDM-5). The corresponding linear matter power spectra have been computed using the formulae provided in \cite{Bode2001} with the damping slope parameter set to $\nu=1$ and the number of degree-of-freedom $g_\textrm{WDM}=1.5$. 
\item LFDM models consisting of realizations with phase transition redshift $z_t=5\cdot 10^5$ (LFDM-1), $8\cdot 10^5$ (LFDM-2) and $15\cdot 10^5$ (LFDM-3). The linear power spectra of these models have been computed with a specifically modified version of {\sc camb} (see \cite{Das2011}). 
\item ULADM models consisting of realizations with particle mass $m_\textrm{a}=1.56\cdot 10^{-22}$ eV (ULADM-1), $4.16\cdot 10^{-22}$ eV (ULADM-2) and $1.54\cdot 10^{-21}$ eV (ULADM-3). We have computed the corresponding linear power spectra with the publicly available code \sc{axionCAMB}\footnote{http://github.com/dgrin1/axionCAMB} \cite{Hlozek2016}.
\end{itemize}

\begin{figure}
\includegraphics[scale=0.4]{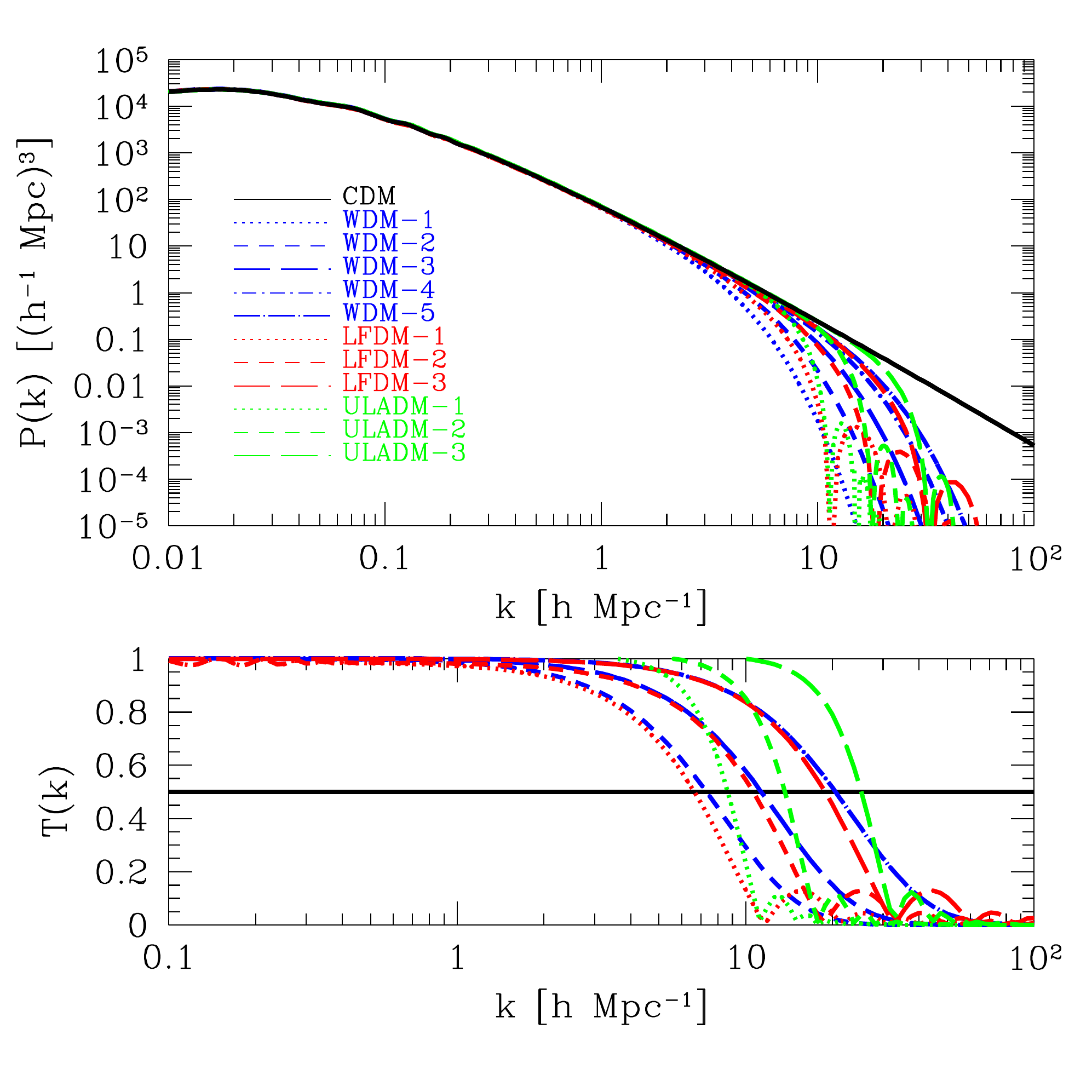}
\caption{Top panel: Linear matter power spectra at $z=0$ for CDM (black solid line), WDM (blue lines), LFDM (red lines) and ULADM models (green lines). Bottom panel: Transfer function of the cut-off matched DM models.}\label{fig1}
\end{figure}

In the top panel of Fig.~\ref{fig1} we plot the linear matter power spectra of the simulated models at $z=0$, while in the bottom panel we plot the transfer functions of the DM models with similar cut-off scale. We can see that the spectra converge to the reference CDM model on the large scales $k\lesssim 1$ h Mpc$^{-1}$, while differences arise in the suppression of power at smaller scales. 

It is worth noticing that with our choice of model parameters, LFDM-1, LFDM-2 and LFDM-3, and ULADM-1, ULADM-2 and ULADM-3 are characterised by power spectra which have a cut-off scale nearly identical to that of WDM-2, WDM-3 and WDM-5 respectively. As can be seen in the bottom panel of Fig.~\ref{fig1}, the corresponding transfer functions are also characterised by very similar half-modes\footnote{The half-mode $k_{1/2}$ is defined has the wavenumber at which the transfer function of a given DM model is half that of the corresponding CDM one \cite{2016PhR...643....1M}: 
\begin{equation}
T(k_{1/2})\equiv\sqrt{\frac{P(k_{1/2})}{P_{\textrm{CDM}}(k_{1/2})}}= \frac{1}{2}.
\end{equation}
Notice that such a definition differs from that commonly used in the literature which defines $k_{1/2}$ as the wavenumber at which the power spectrum of a given DM model is half the value of the CDM one.}, these are quoted in Table~\ref{tab0}. Despite such similarities, we can see that these models exhibit a different distribution of power for $k\gtrsim k_{1/2}$. Hence, it is reasonable to expect that this may lead to differences in the high-redshift abundance of small-mass halos for these models, a point which we discuss in detail in Section~\ref{hmf}.

\begin{table}
\centering
\begin{tabular}{|c|c|c|c|c|c|}
\hline\hline
Model & $k_{1/2}$ &Model & $k_{1/2}$  & Model & $k_{1/2}$ \\
\hline
WDM-2 &   7.3  &  LFDM-1  & 6.6   & ULADM-1 & 8.7\\
\hline
WDM-3 &   11.3 & LFDM-2  & 10.6 & ULADM-2 & 13.7\\
\hline
WDM-5 &   20.4 & LFDM-3  & 18.8 & ULADM-3 & 25.2\\
\hline\hline
\end{tabular}
\caption{\label{tab0}Half-mode values in units of h Mpc$^{-1}$.}
\end{table}

\subsection{Simulation characteristics}\label{simuchar}
In principle, the use of N-body methods to simulate the non-linear structure formation of the non-standard DM cosmologies described above may not be a valid approach. This is because these models are characterised by microphysical processes that distinguish their particle dynamics from that of a purely collisionless DM component.

As an example, in the case of WDM models, one should in principle account for the distribution of thermal velocities \cite{Bode2001,Lowell2014}. This effect is usually implemented in numerical simulations as a random kick applied to the N-body particles (which trace the clustering of the matter density field), even though the root-mean-square velocity of WDM particles is several order of magnitudes smaller than that arising during the non-linear gravitational collapse. However, as pointed out in \cite{Angulo2013}, N-body particles are a coarse-grained representation of the phase-space distribution of the microscopic particles. Therefore the addition of a kick is equivalent to inducing a local coherent motion of a large ensemble of microscopic WDM particles, which leads to a velocity spectrum that is inconsistent with results from linear perturbation theory \cite{Colin2008}. In the case of fermionic WDM the Tremaine-Gunn effect \cite{Tremaine1979} leads to modified halo density profiles, however this occurs on scales smaller than those probed by the LF. Thus, as we are interested in deriving the mass distribution of DM halos, we can safely neglect this effect and run N-body simulations of WDM models with appropriate initial power spectra.

\begin{figure}
\includegraphics[scale=0.4]{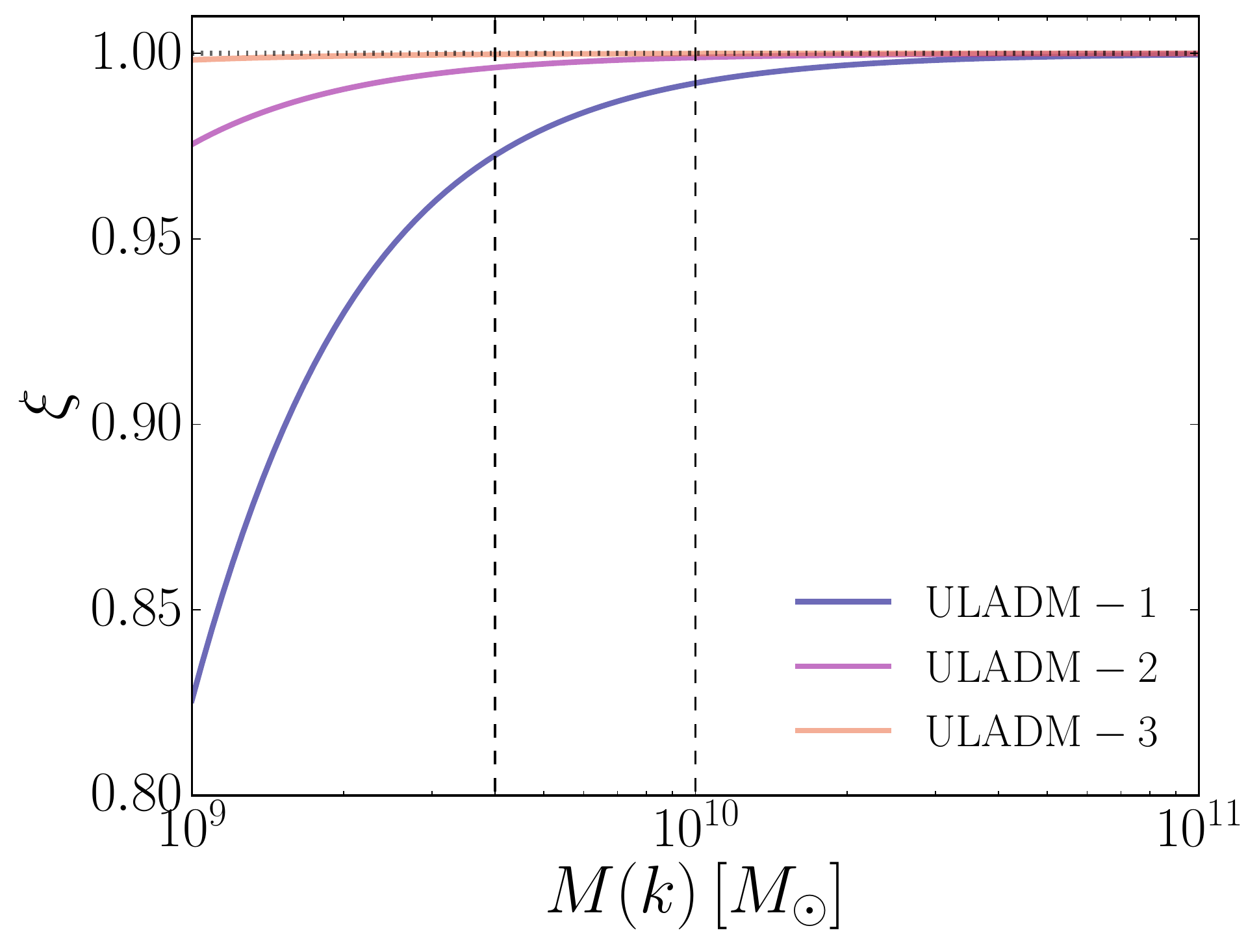}
\caption{The growth rate ratio, $\xi$, defined in \cite{Schive2016} for the ULADM models for $z_{\rm start}=100$ and $z_{\rm end}=4$ as function of the mass scale $M(k)=4\pi(\pi/k)^3\rho_m/3$, where $\rho_m$ is the comoving matter density. The vertical dotted lines denotes the lowest mass scales probed by \textit{HST} observations of the faintest galaxies at $4\lesssim z\lesssim 8$. Unlike \cite{Schive2016}, our initial conditions use the correctly evolved initial conditions from \textsc{axionCAMB}; we use $\xi$ as a measure of the importance of scale-dependent growth to assess how well ULAs can be modelled with $N$-body simulations. The lightest ULA model we consider is well described by collisionless simulations at the percent level for halo masses of interest.}\label{fig_xi}
\end{figure}

In the case of ULADM models, realistic simulations should solve the coupled Schr{\"o}dinger-Poisson system (see e.g. \cite{Schive2014,Schwabe2016} or \cite{Veltmaat2016} for the particle in cell approach) to account for quantum wave-like effects that are specific to this class of models. However, as in the study by Schive et al. \cite{Schive2016}, the mass scales and redshifts which are of interest to our analysis are mostly insensitive to these effects. This is demonstrated in Fig.~\ref{fig_xi}, where we plot the growth rate ratio, $\xi$, defined in \cite{Schive2016} using the exact growth solution in \cite{2016PhR...643....1M}. As we can see, over the mass scale interval probed by LF observations the scale dependent growth rate of our lightest ULADM model differs from the CDM case by $\lesssim 5\%$. Thus, for our purposes we can safely simulate the non-linear clustering of these models using the N-body method.

The use of N-body simulations is also justified in the case of the LFDM models, since in this scenario DM particles become collisionless soon after matter-radiation equality. Thus, by the initial redshift of the simulations the system becomes practically collisionless and its non-linear clustering can be followed through the dynamics of N-body particles with the appropriate initial power spectrum. 
 
We run the code {\sc ramses} \cite{Teyssier2002} to perform a series of high-resolution N-body simulations with the goal of resolving the low-mass end of the high-redshift halo mass function for the DM models described above. To this purpose we have simulated ($27.5\,h^{-1}$ Mpc)$^3$ volumes with $N_p=1024^3$ particles corresponding to a particle mass resolution $m_p=1.61\cdot 10^6\,{\rm M_\odot}\,h^{-1}$. 

\begin{figure*}[ht]
\includegraphics[scale=0.53]{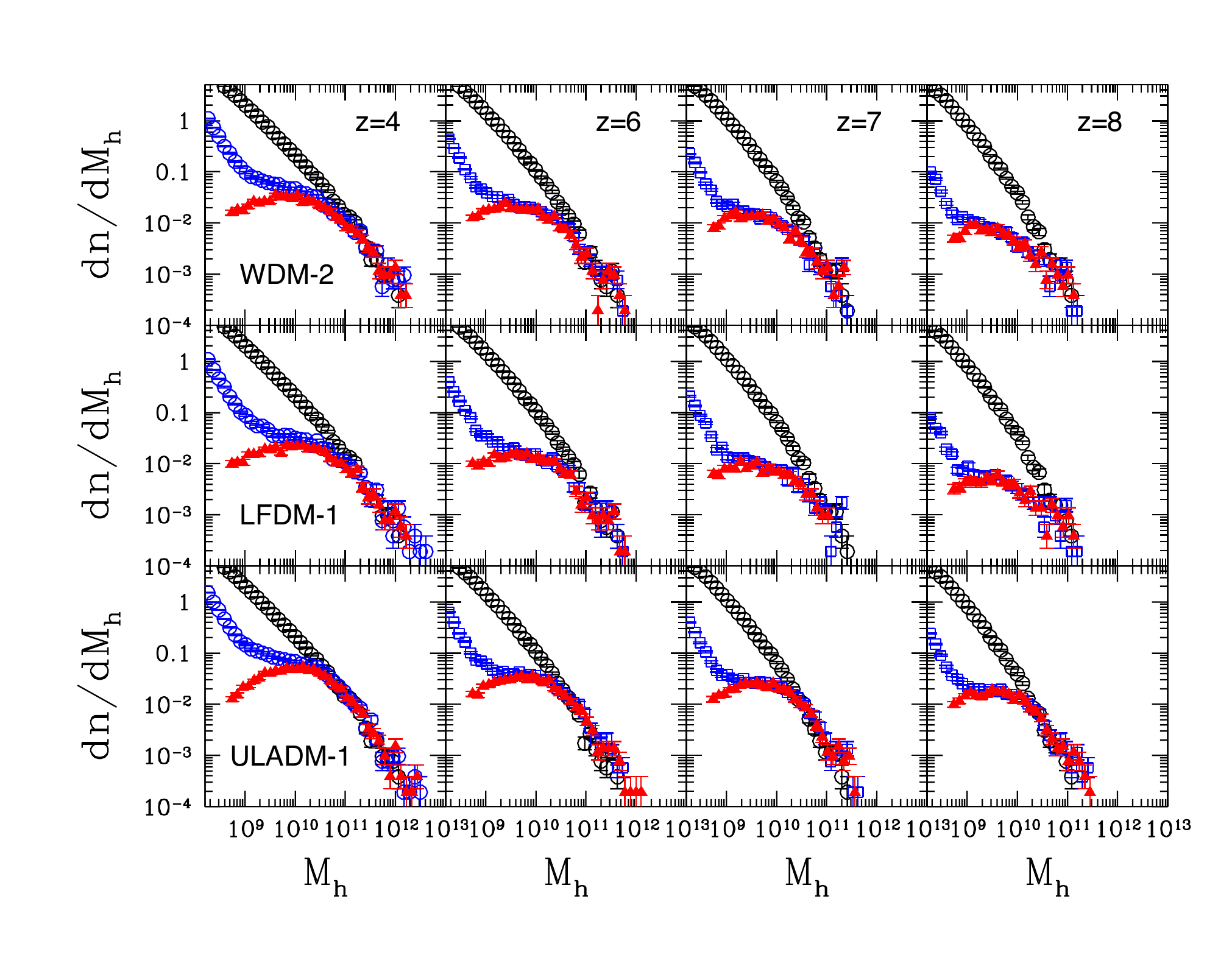}
\caption{Halo mass function at $z=4,6,7$ and $8$ (panels left to right) in mass bins of size $\Delta{\rm M_h}/{\rm M_h}=0.20$ for WDM-2, LFDM-1 and ULADM-1 models (panels top to bottom respectively) before (blue empty squares) and after (red filled triangles) halo selection. In each panel the CDM halo mass function is shown as empty circles. Error-bars correspond to the Poisson errors in each mass bin.}\label{fig2}
\end{figure*}

We generate initial conditions using the Zel'dovich approximation as implemented in {\sc mpgrafic} \cite{Prunet2008}. For all models we use the same phase of the initial conditions and set the starting redshift $z_i$ of the simulations such that for a given model the standard deviation of the linear density field smoothed on the scale of the coarse grid is given by $\sigma(\Delta_x^{\rm coarse},z_i)=0.02$. Enforcing this constraint gives sufficiently high initial redshifts such as to guarantee that deviations from the Zel'dovich approximation remain negligible. For each model simulation we store eleven snapshots between $z=0$ and $10$.

The simulations were run on the CURIE supercomputer of the Institute for Development and Resources in Intensive Scientific Computing (IDRIS) using 1024 processors for a total running time of 2 million hours.

\subsection{Halo finder and spurious halo selection}
We detect halos using a parallelised version of the friend-of-friend algorithm \cite{Davis1985} implemented in the code pFoF \cite{Roy2014}. This identifies halos as group of particles with a given linking length parameter $b$, which we set to $b=0.2$. 

In order to reduce the impact of mass resolution errors, one may conservatively consider halos with at least 100 particles. However, in the case of cosmological models with suppressed spectra at small scales, the sampling of Poisson noise between the cut-off scale of the power spectrum and the Nyquist frequency of the simulations leads to the formation of spurious numerical halos, which cause an unphysical upturn of the halo mass function at low masses \cite{Gotz2002,Gotz2003,WangWhite2007}. 

Several empirical methods have been investigated in the literature to identify and remove artificial halos. For instance, Wang \& White \cite{WangWhite2007} have suggested to cut halo catalogs below a mass limit ${\rm M_{lim}}=10.1\bar{\rho}\,d\,k_{\rm peak}^{-2}$ where $\bar{\rho}$ is the mean matter density, $d$ is the mean intra-particle distance of the simulation, $k_{\rm peak}$ is the location of the peak in the dimensionless linear power spectrum $\Delta^2(k) \equiv k^{3}P(k)/2\pi^2$ and the numerical coefficient is estimated from simulations. A more sophisticated approach has been proposed in \cite{Lowell2014}, which relies on the idea that genuine proto-halos are spheroidal, thus spurious halos are identified as groups of particles associated to Lagrangian patches in the initial conditions characterised by a shape flatter than a certain threshold. 

Here, we follow the approach presented in \cite{Agarwal2015}. This builds upon the fact that the structural properties of halos as described by the spin and shape parameters as well as the degree of relaxation provide distinct physical proxies of the genuine nature of halos in the simulations. In \cite{Agarwal2015}, the analysis of halo catalogs from WDM and LFDM simulations has shown that halos contributing to the upturn in the halo mass function are characterised by highly distorted statistical distributions of halo spin and shape parameters, which strongly correlate with large deviations from the virial condition as measured by the parameter $\eta\equiv 2K/|E|$,\footnote{It is worth reminding that in the case of ULADM models the virial theorem is modified by the presence of the quantum pressure $Q$ such that $\eta=2K/(|E|+2Q)$ \cite{Schive2014b,Hui2016}. This is not taken into account in our simulations, and we leave a detailed study to future work.} where $K$ is the total kinetic energy of the halo particles and $|E|$ its gravitational potential energy. 

As shown in \cite{Agarwal2015}, retaining halos with at least 300 particles and with deviations from the virial state in the range $0<\eta<1.5$ are sufficient conditions to remove the bulk of spurious objects from the numerical halo catalogs and recover undistorted statistical distributions for halo spin and shape parameters independently of the mass resolution of the simulations. 

To illustrate the effect of spurious halos in our simulation suite we plot in Fig.~\ref{fig2} the halo mass function at $z=4,6,7$ and $8$ (panels from left to right) for WDM-2, LFDM-1 and ULADM-1 models (panels from top to bottom respectively). In each panel the black points represent the CDM mass function, while blue (red) points denote the non-standard DM model mass function before (after) spurious halo selection. Error-bars are given by the Poisson error within each mass bin. For each model we can see that at the high-mass end the mass function converges to that of CDM. This is expected, since on large scales the models have linear matter power spectra identical to that of the CDM case. At low masses (${\rm M_h}\lesssim 10^{10}{\rm M_\odot}\,h^{-1}$) we can see the characteristic upturn which is indicative of the presence of spurious halos. In particular, the lower the redshift the higher the amplitude of the upturn, consistent with the expectation that the number of spurious halos increases as the simulation evolves from earlier to later times. After halo selection, the upturn disappears and we recover the expected halo abundance suppression at the low-mass end. 

We apply the selection criteria of \cite{Agarwal2015} to all numerical halo catalogs. Since we consider halos with no less than 300 particles, the selected halo catalogs have a mass cut at ${\rm M^{min}_h}\approx 5\cdot 10^8$ M$_\odot\,h^{-1}$. Therefore, with the intent of being as conservative as possible we will not extrapolate any result below ${\rm M^{min}_h}$.

\subsection{Halo mass function}\label{hmf}

\begin{table}
\centering
\begin{tabular}{|c|c|c|c|}
\hline\hline
z & A & a & p\\
\hline
4 &   0.35620   &  0.94020   &  -0.87256 \\
\hline
5 &   0.29542   &  0.94630   &  -0.99538 \\
\hline
6 &   0.29188   &  0.82990   &  -0.84125 \\
\hline
7 &   0.21785   &  0.91178   &  -1.1072    \\
\hline
8 &   0.25823   &  0.79364   &  -0.81431  \\
\hline\hline
\end{tabular}
\caption{\label{tab1}Sheth-Tormen multiplicity function best-fit coefficients to the CDM mass function at $4\le z\le 8$.}
\end{table}

\begin{figure}
\includegraphics[scale=0.4]{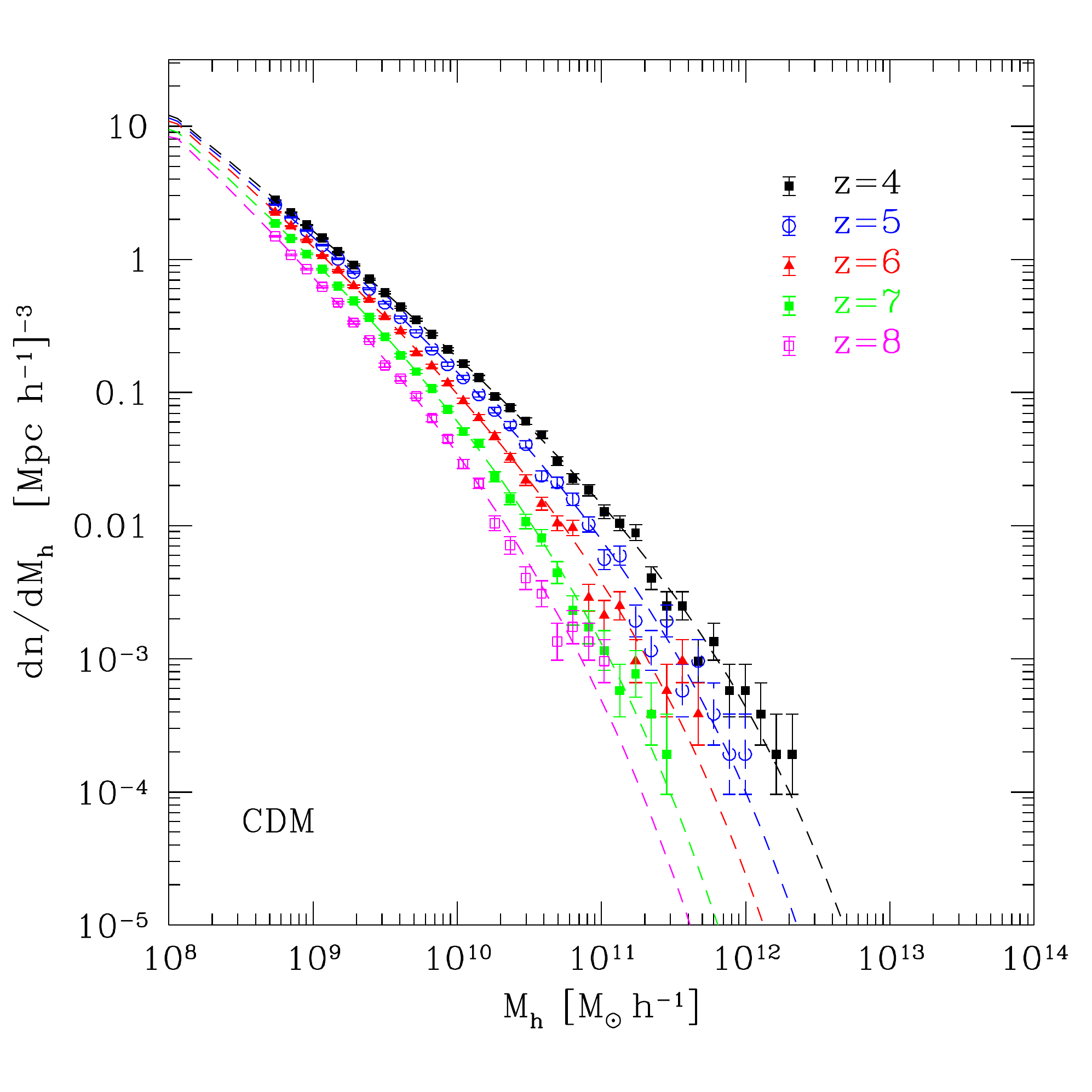}
\caption{Halo mass function of the reference CDM model simulation at $z=4,5,6,7$ and $8$ in mass bins of size $\Delta{\rm M_h}/{\rm M_h}=0.20$. The dashed lines corresponds to Sheth-Tormen multiplicity function, Eq.~(\ref{CDMfit}), with best-fit parameters at the different redshifts given in  Table~\ref{tab1}.}\label{fig3}
\end{figure}

We fit the halo mass function of the CDM model as:
\begin{equation}\label{CDMfit}
\frac{dn}{d{\rm M_h}}\Bigr|_{{\rm CDM}}=\frac{\bar{\rho}}{{\rm M_h}^2}\frac{d\ln{\sigma^{-1}}}{d\ln{{\rm M_h}}}f_{\rm ST}(\delta_c/\sigma),
\end{equation}  
where $\bar{\rho}$ is the mean matter density, $\sigma$ is the root-mean-square of the linear density field smoothed on the scale enclosing a spherical volume of mass ${\rm M_h}$, and $f_{\rm ST}(\delta_c/\sigma)$ is the Sheth-Tormen (ST) multiplicity function \cite{Sheth1999}:
\begin{equation}
f_{\rm ST}(\nu)=2\,A\sqrt{\frac{a\nu}{2\pi}}e^{-\frac{a\nu}{2}}\left[1+\frac{1}{(a\nu)^p}\right],
\end{equation}
where $\nu=(\delta_c/\sigma)^2$ with $\delta_c$ being the linear spherical collapse threshold computed using the formula given in \cite{Kitayama1996}. We determine the best-fit ST coefficients using a Levenberg-Marquardt minimisation scheme. These are quoted in Table~\ref{tab1} for the redshifts of interest. In Fig.~\ref{fig3} we plot the numerical mass functions against the ST best-fits.

We use the CDM calibrated formula to fit the mass function of the non-standard DM simulations using the following parameterisation:
\begin{equation}
\frac{dn}{d{\rm M_h}}=10^{\alpha+\beta\frac{{\rm M_*}}{{\rm M_h}}}\left(1-e^{-\frac{\rm M_h}{\rm M_*}}\right)^{\gamma}\frac{dn}{d{\rm M_h}}\Bigr|_{{\rm CDM}},\label{dndm_ndm}
\end{equation}
where $\alpha$, $\beta$, $\gamma$ and ${\rm M_*}$ are parameters which we best-fit against the N-body mass function. We prefer to work with such a parameterisation rather than the formula introduced in \cite{Angulo2013}, since it provides better fits to the numerical data. Accordingly, we find the best-fit functions to have reduced $\chi_{\rm red}^2\approx 1$. In Appendix~\ref{app1} we illustrate the goodness-of-fit of Eq.~(\ref{dndm_ndm}) and quote the values of the best-fit coefficients for all simulated DM models.\footnote{In order to avoid confusion we want to remark that the identification of ${\rm M_*}$ as a mass scale cut-off related to a DM model parameter such as the thermal relic mass in the case of WDM models or the axion mass in the case of ULADM models is valid only for $\beta=0$. Since we find $\beta\ne 0$ for all models, one should not associate any particular physical meaning to the redshift evolution of the best-fit values of ${\rm M_*}$ quoted in Tables~\ref{tabA1},~\ref{tabA2} and~\ref{tabA3}.}

\begin{figure*}
\includegraphics[scale=0.6]{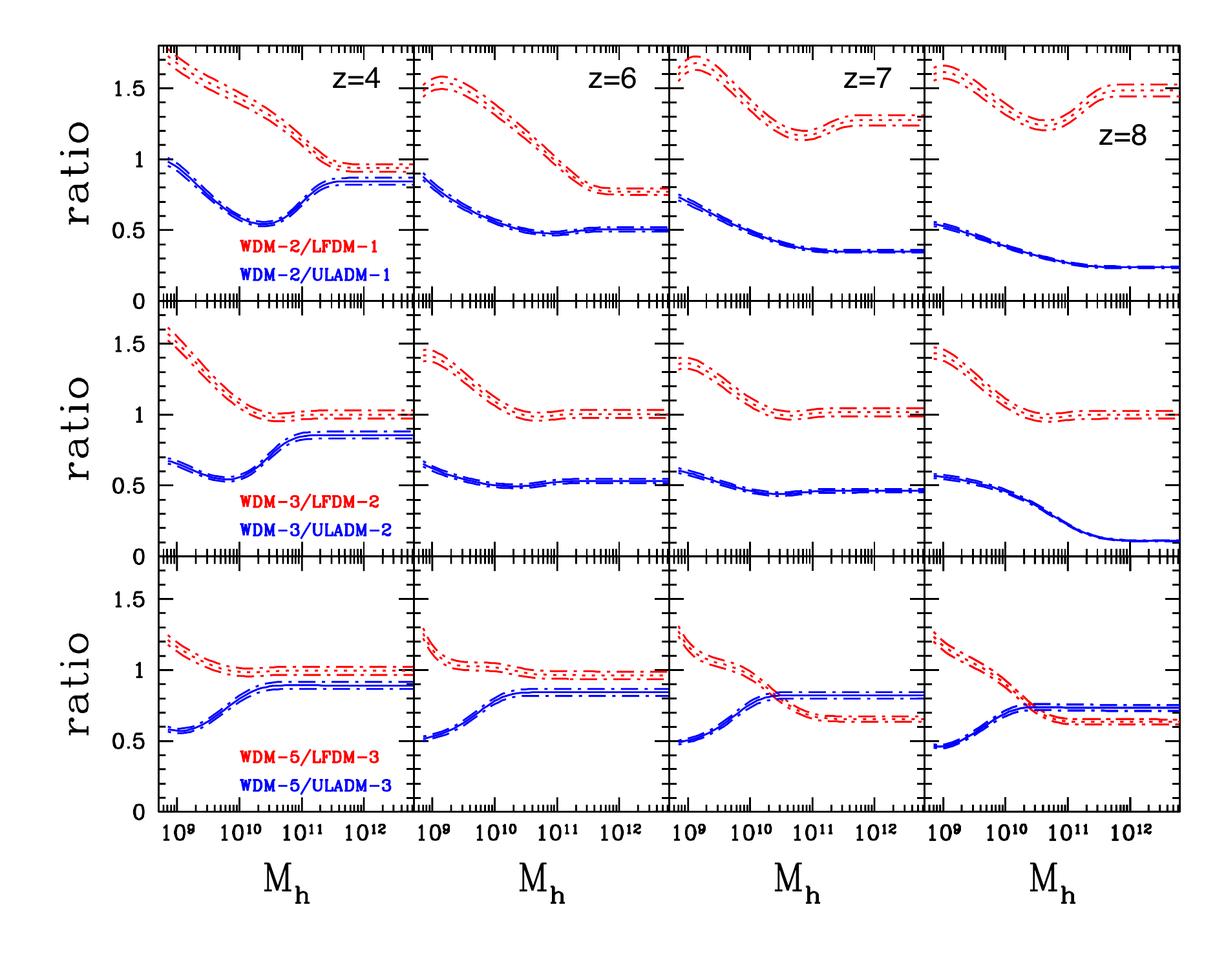}
\caption{Ratio of the best-fit mass functions at $z=4,6,7$ and $8$ (left to right panels respectively) for WDM-2/LFDM-1 and WDM-2/ULADM-1 (top panels), WDM-3/LFDM-2 and WDM-3/ULADM-2 (central panels), WDM-5/LFDM-3 and WDM-5/ULADM-3 (bottom panels). Ratios with respect to the LFDM are given by the red dotted lines, while those relative to the ULADM models are shown as blue solid lines. The short-dash-dot lines above and below each curve given the numerical statistical errors around the ratio of the best-fit functions.}\label{fig3bis}
\end{figure*}

As discussed in Section~\ref{simuchar} the models LFDM-1, LFDM-2 and LFDM-3, and ULADM-1, ULADM-2 and ULADM-3 are characterised by a power spectrum cut-off which is nearly identical to that WDM-2, WDM-3 and WDM-5 respectively. Nevertheless, we find differences among these models for the predicted abundance of low-mass halos at high-redshifts which are well above the numerical statistical uncertainties ($\sim 2\%$ level). This can be seen in Fig.~\ref{fig3bis}, where we plot the ratio of the best-fit mass function at $z=4,6,7$ and $8$ (left to right panels) for WDM-2/LFDM-1 and WDM-2/ULADM-1 (top panels), WDM-3/LFDM-2 and WDM-3/ULADM-2 (central panels), and WDM-5/LFDM-3 and WDM-5/ULADM-3 (bottom panels). Ratios with respect to the LFDM models are shown as red dotted lines, while those with respect to ULADM models are shown as blue solid lines. In particular, we can see that the WDM models have systematically greater abundances of low-mass halos relative to the LFDM counterparts, while the opposite occurs for the ULADM modes. As can be seen in the lower panel of Fig.~\ref{fig1}, this is consistent with the fact that the LFDM models have transfer functions that for $k\gtrsim k_{1/2}$ are systematically lower than the WDM ones, while the ULADM transfer functions are larger. We will see that these differences manifest in different predictions of the faint-end slope of the LF at high-redshifts.

\section{Methodology}\label{method}
Our goal is to constrain the simulated DM models using an up-to-date compilation of high-redshift LF data at $z=6,7$ and $8$ covering an unprecedented UV-magnitude range from the brightest to the very faint. In order to convert the N-body calibrated mass functions into LF model predictions that can be compared to the data it is necessary to specify a relation between halo mass and UV-luminosity. As already mentioned in Section~\ref{intro}, past studies in the literature computed such a relation using HAM methods \cite{Schultz2014,Bozek2015} or assumed a parametric form to be constrained by the LF data \cite{Schive2016}. 

Here, we adopt a hybrid method with the intent of gaining insight on the evolution of the star formation rate of high-redshift galaxies as a function of redshift and host halo mass, as well as assessing the impact of dust extinction on the rest-frame UV-luminosities. The approach used here can be summarized as follows:
\begin{itemize}
\item LF measurements are corrected for dust extinction using the established relation between extinction $A_{UV}$ and UV-continuum slope $\beta_{UV}$ from \cite{Meurer1999}. In particular, we calibrate an extinction model using UV-continuum slope measurements from Bouwens et al. \cite{Bouwens2014} to correct LF data at $z=4$ and $5$ from \cite{Bouwens2015}. The implementation of this extinction correction is described in Section~\ref{dustext};
\item the corrected LF data at $z=4$ and $5$ is converted into the respective SFR densities $\varphi({\rm SFR})$ using the Kennicutt-relation \cite{Kennicutt1998}.
\item The inferred $\varphi({\rm SFR})$ functions at $z=4$ and $5$, together with the halo mass function fits to the N-body simulations, are used to derive ${\rm SFR}({\rm M_h})$ relation using HAM technique. The ${\rm SFR}({\rm M_h})$ relations at $z=4$ and $5$ are then redshift-averaged to obtain the average relation, ${\rm SFR_{av}}({\rm M_h})$. This is repeated for each of the simulated DM model. This step is described in Section~\ref{sfrav};
\item For each DM model, after converting the SFR back into UV-luminosities we model the LF at $z=6,7$ and $8$ by integrating over the halo mass function a log-normal SFR probability density distribution with average $\langle {\rm SFR(M_h)}\rangle \equiv \varepsilon^z_{\rm SFR} {\rm SFR_{av}}({\rm M_h})$ and intrinsic dispersion $\sigma^z_{\varepsilon_{\rm SFR}}$, where $\varepsilon^z_{\rm SFR}$ and $\sigma^z_{\varepsilon_{\rm SFR}}$ are free parameters. This will be explained in detail in Section~\ref{LFmodel};
\item Finally, a $\chi^2$-analysis of LF data at $z=6,7$ and $8$ is performed to determine the best-fit values of $\varepsilon^z_{\rm SFR}$ and $\sigma^z_{\varepsilon_{\rm SFR}}$, and evaluate the goodness-of-fit for each DM model.
\end{itemize}

To avoid confusion, hereafter we refer to density function as $\varphi$ to denote $\varphi \equiv d\Phi / d\log_{10}X$, and as $\phi$ to denote $\phi \equiv d\Phi/dX$, where $\Phi$ is the cumulative density function.

\subsection{Dust extinction correction \& SFR densities}\label{dustext}
We correct the LF data using the empirical relation between extinction and UV-continuum slope, $A_{UV}=4.43+1.99\,\beta_{UV}$ \cite{Meurer1999}. Following \cite{Smit2012}, at each $\rm M_{UV}$ we assume $\beta_{UV}$ to be normal distributed with mean $\langle\beta_{UV}\rangle$ and dispersion $\sigma_{\beta_{UV}}=0.34$, giving the average extinction $\langle A_{UV}\rangle = 4.43+0.79 \ln{(10)}\,\sigma_{\beta_{UV}}^2+1.99\,\langle\beta_{UV}\rangle$. $A_{UV}$ is set to zero when $A_{UV}<0$.

We model the mean slope $\langle\beta^z_{UV}\rangle$  as in \cite{Trenti2015,Mason2015}:
\begin{equation}
\begin{multlined}
\langle\beta^z_{UV}\rangle=\\ 
\begin{cases}
\left[\beta^z_{M_0}-c\right]\exp{\left[\frac{d\beta^z}{dM_0}\cdot\frac{{\rm M_{UV}}-M_0}{\beta^z_{M_0}-c}\right]}+c,\,\,\, {\rm M_{UV}}\ge M_0\\
\frac{d\beta^z}{dM_0}\cdot\left[{\rm M_{UV}}-M_0\right]+\beta^z_{M_0},\,\,\,\,\,\,\,\,\,\,\,\,\,\,\,\,\,\,\,\,\,\,\,\,\,\,\,\, {\rm M_{UV}}<M_0
\end{cases}
\end{multlined}
\end{equation}
where $c=-2.33$ and $M_0=-19.5$. We approximate $\beta^z_{M_0}$ and $d\beta^z/dM_0$ as linear redshift functions with coefficients determined by least-square interpolation of the values given in Table 3 of \cite{Bouwens2014}. This gives:
\begin{equation}
\begin{cases}
\beta^z_{M_0}=-1.573630 - 0.069756 \cdot z,\\
\frac{d\beta^z}{dM_{0}}= -0.095379 - 0.006827 \cdot z .
\end{cases}
\end{equation}

The extinction correction has a twofold effect on the LF function (see e.g. \cite{Smit2012}). First, it shifts the LF toward brighter magnitudes since
\begin{equation}\label{muvc}
{\rm M_{UV}^c}={\rm M_{UV}}-\langle A_{UV}({\rm M_{UV}},z)\rangle,
\end{equation}
where $\langle A_{UV}({\rm M_{UV}},z)\rangle \ge 0$. Second, it alters the magnitude bin-size and consequently the amplitude of the LF. Denoting with $\Delta{\rm M_{UV}}$ the bin-size of the uncorrected LF, we have
\begin{eqnarray}
\Delta {\rm M_{UV}^c}&=& \Delta {\rm M_{UV}} -\left\langle A_{UV}\left({\rm M_{UV}}+\frac{\Delta {\rm M}}{2},z\right)\right\rangle+\nonumber\\
&+&\left\langle A_{UV}\left({\rm M_{UV}}-\frac{\Delta {\rm M}}{2},z\right)\right\rangle.
\end{eqnarray}
Thus the corrected LF reads as
\begin{equation}\label{phimuvc}
\phi({\rm M_{UV}^c}) = \phi({\rm M_{UV}})\frac{\Delta {\rm M_{UV}}}{\Delta {\rm M_{UV}^c}}.
\end{equation}

\begin{figure}
\includegraphics[scale=0.4]{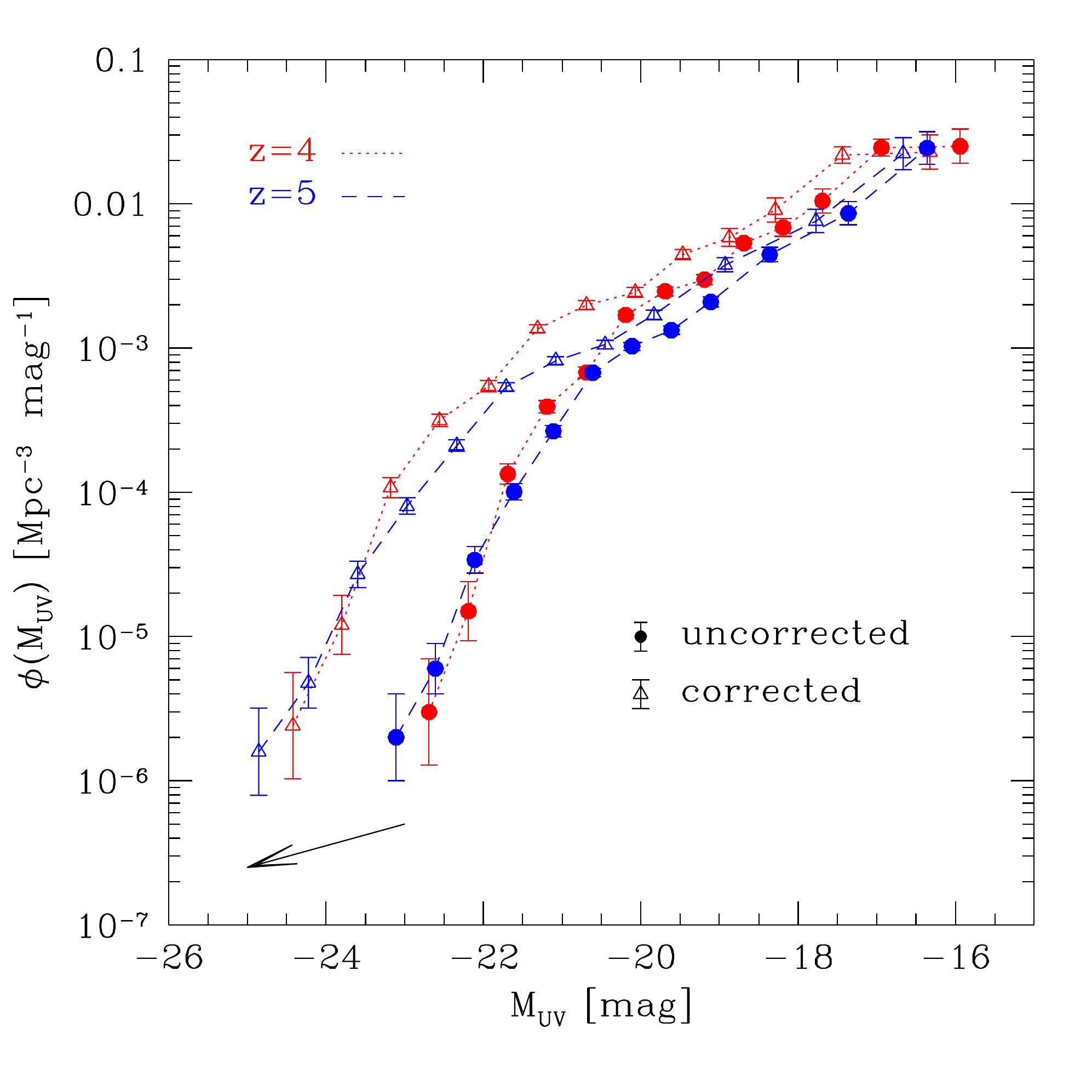}
\caption{Galaxy luminosity function at $z=4$ (red points) and $z=5$ (blue points) from Bouwens et al. \cite{Bouwens2015} before (filled circles) and after (empty triangles) correction for dust extinction. The arrow indicates the direction of the shift of the uncorrected data due to the extinction correction.}\label{fig5}
\end{figure}

We use Eq.~(\ref{muvc}) and Eq.~(\ref{phimuvc}) to correct the LF measurements (including errors) from Bouwens et al. \cite{Bouwens2015} at $z=4$ and $5$. These are displayed in Fig.~\ref{fig5}, where we plot the LF data (including errorbars) at $z=4$ (red points) and $z=5$ (blue points) before (filled circles) and after (empty triangles) correction for dust extinction.

\begin{figure}
\includegraphics[scale=0.4]{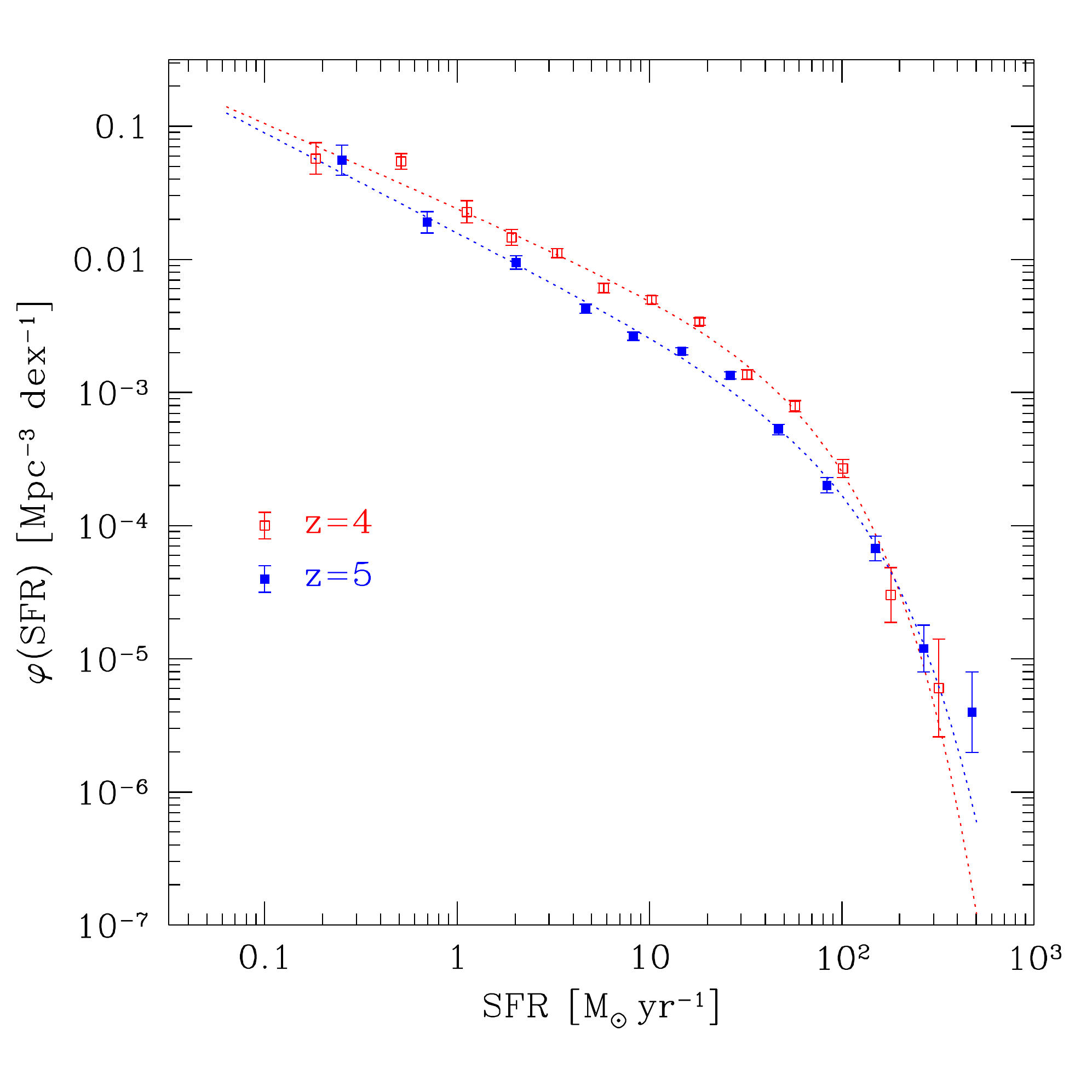}
\caption{SFR density function estimates at $z=4$ (red empty squares) and $z=5$ (blue filled squares) obtained using dust-corrected LF data from Bouwens et al. \cite{Bouwens2015}. The dashed lines represent the Schechter function (Eq.~\ref{sfrf}) with best-fit coefficients (see Table~\ref{tab2}).}\label{fig6}
\end{figure}

We convert the corrected LFs into SFR densities using the Kennicutt-relation ${\rm SFR}\,[{\rm M}_{\odot}\,{\rm yr}^{-1}]=1.25\cdot 10^{-28} L_{UV}\,[{\rm erg}\,{\rm s}^{-1}{\rm Hz}^{-1}]$\footnote{Absolute magnitudes and luminosities in {\it cgs} units are related as:
\begin{equation}
\frac{L_{UV}}{{\rm erg}\,{\rm s}^{-1}{\rm Hz}^{-1}}=4\pi\,(10\cdot3.8856802\cdot10^{18})^2\times 10^{-\frac{{\rm M_{UV}}+48.6}{2.5}}
\end{equation}
} \cite{Kennicutt1998}. We fit the SFR density estimates with a Schechter fitting function:
\begin{equation}\label{sfrf}
\varphi({\rm SFR})=\ln(10) \,\varphi_{\rm SFR}^*\left(\frac{\rm SFR}{{\rm SFR}_*}\right)^{\alpha_{\rm SFR}+1}e^{-\frac{\rm SFR}{{\rm SFR}_*}}.
\end{equation}
The best-fit values of $\varphi_{\rm SFR}^*$, $\alpha_{\rm SFR}$ and ${\rm SFR}_*$ at $z=4$ and $5$ are quoted
in Table~\ref{tab2}, while in Fig.~\ref{fig6} we plot the SFR density function measurements against the best-fit Schechter functions.

\begin{table}[t]
\centering
\begin{tabular}{|c|c|c|c|}
\hline\hline
z & $\varphi_{\rm SFR}^*$ [Mpc$^{-3}$ dex$^{-1}$]& $\alpha_{\rm SFR}$ & SFR$_*$ [M$_\odot$ yr$^{-1}$]\\
\hline
4 &  0.49055E-03   &  -0.16551E+01 &  0.46798E+02\\
\hline
5 &  0.29502E-03   & -0.16995E+01 & 0.47624E+02\\
\hline\hline
\end{tabular}
\caption{\label{tab2} Schechter function (Eq.~\ref{sfrf}) best-fit coefficients to SFR densities at $z=4$ and $5$.}
\end{table}

\subsection{Average ${\rm M_h}$-SFR relation}\label{sfrav} 
\begin{figure*}[ht]
\centering
\subfigure{\includegraphics[scale=0.4]{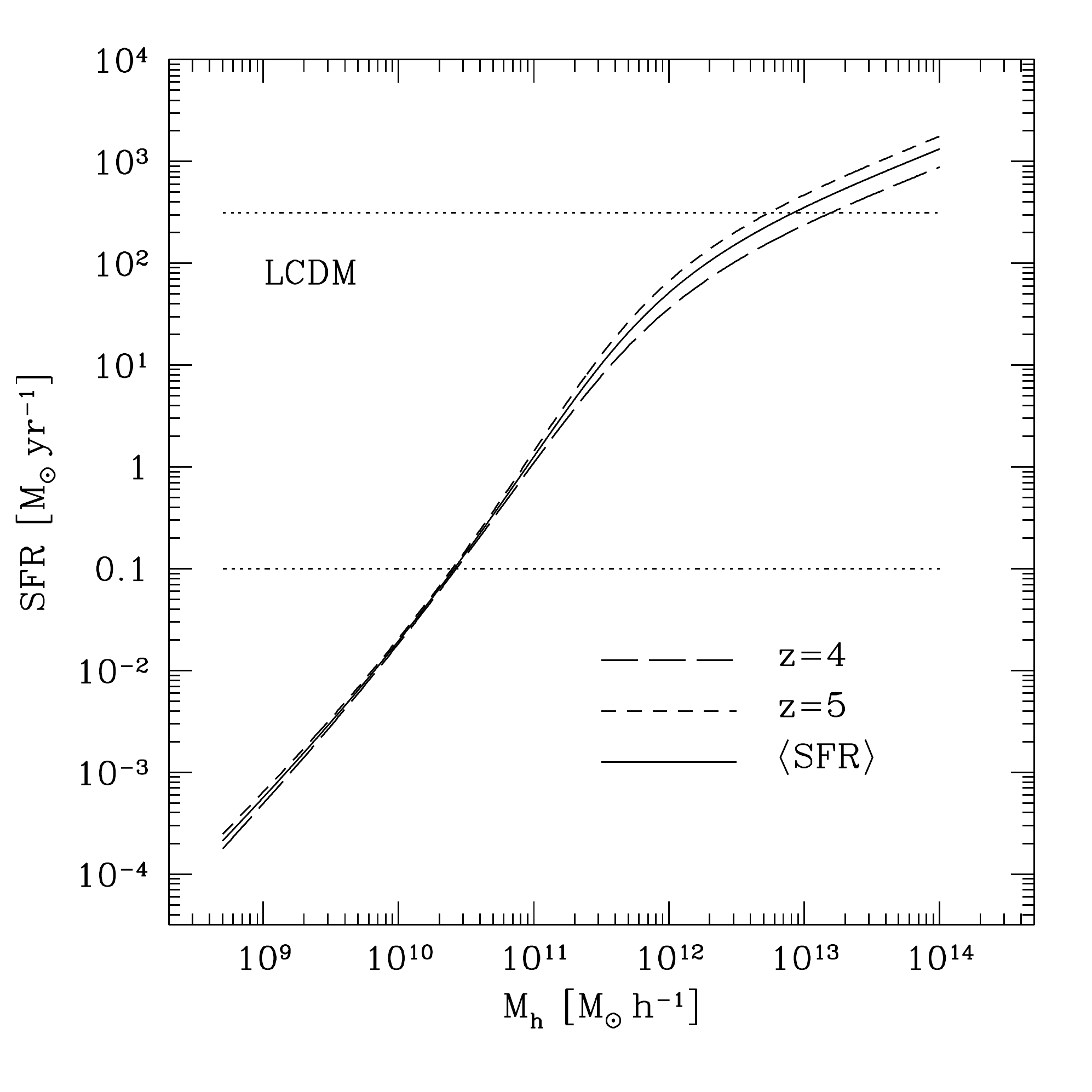}}\quad
\subfigure{\includegraphics[scale=0.4]{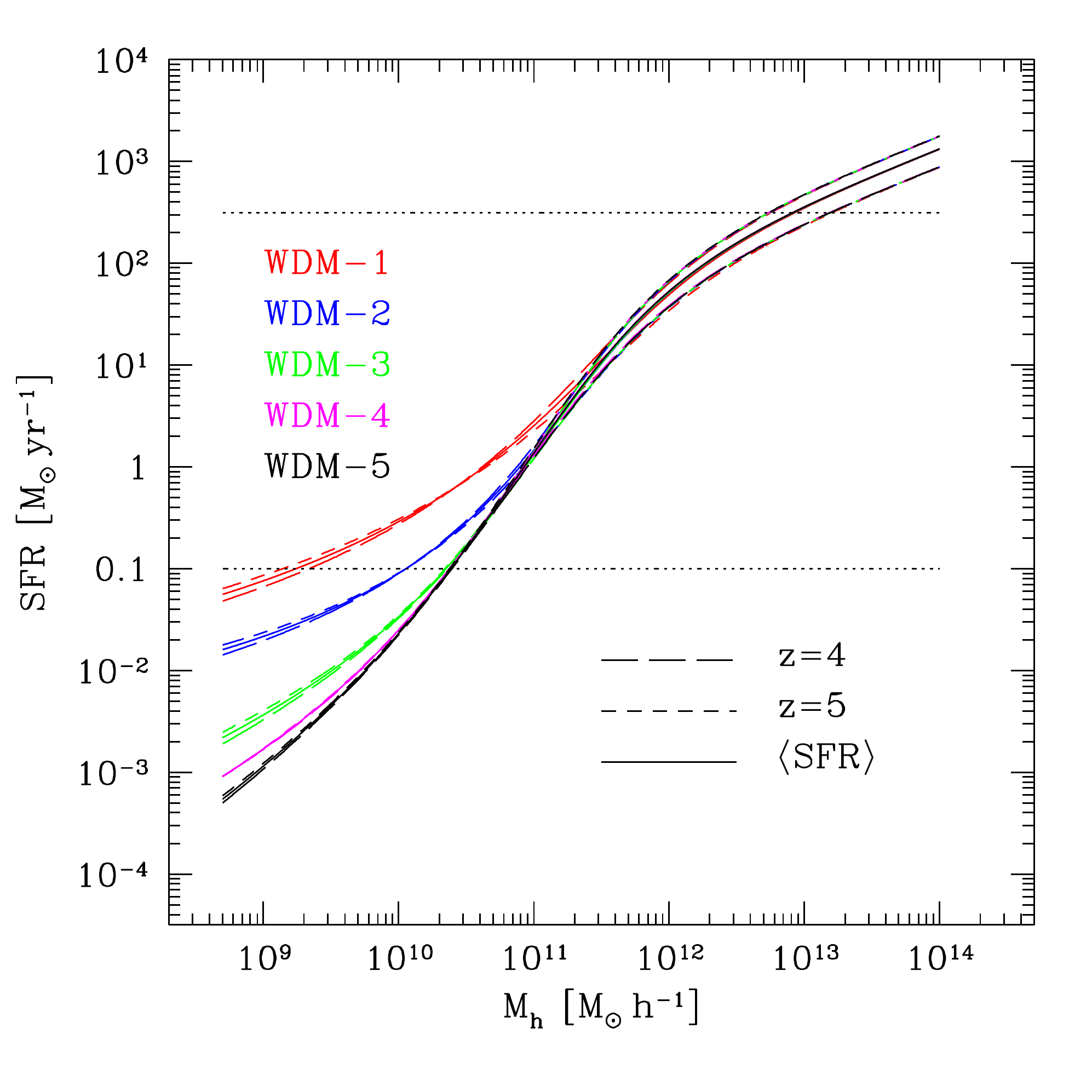}}\quad
\subfigure{\includegraphics[scale=0.4]{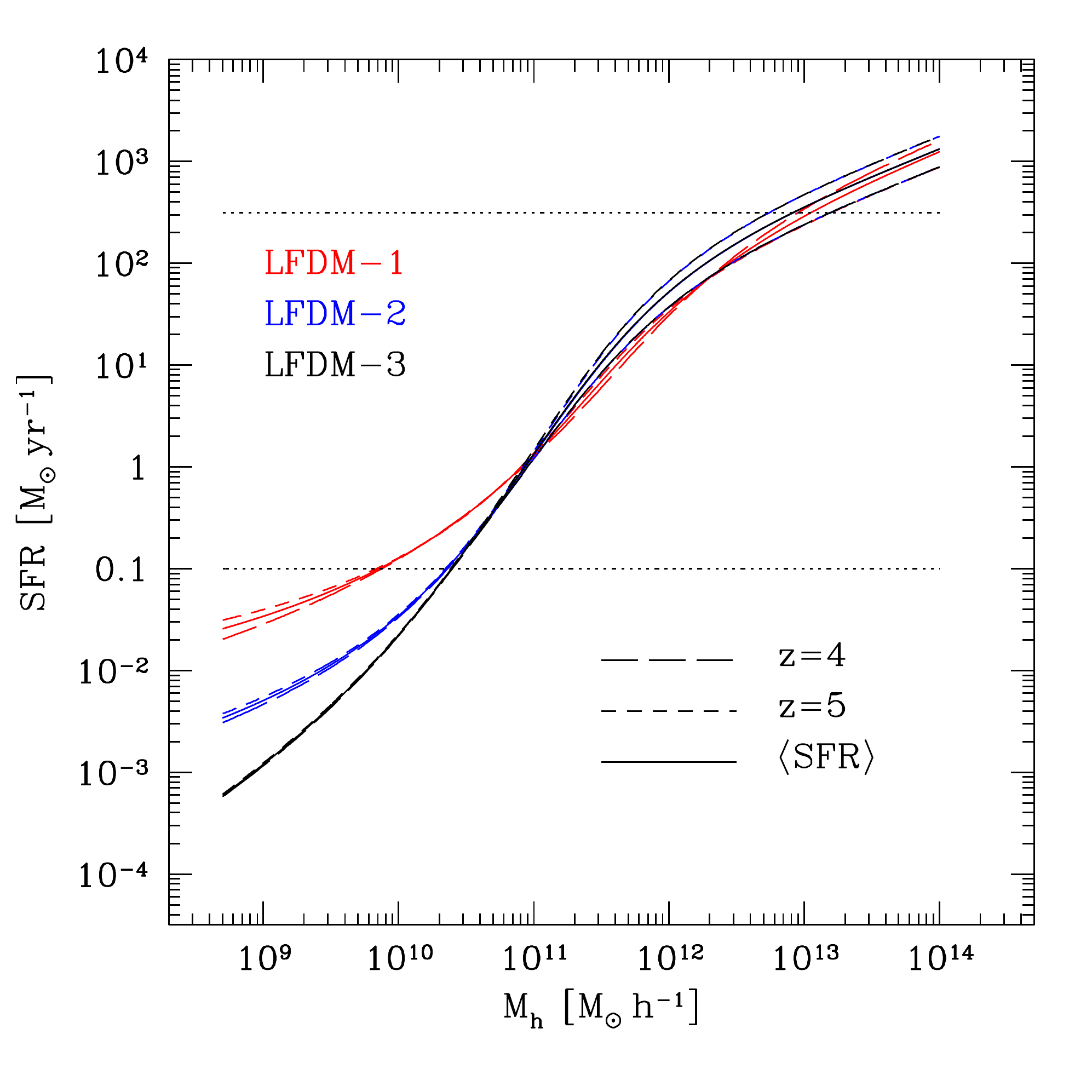}}\quad
\subfigure{\includegraphics[scale=0.4]{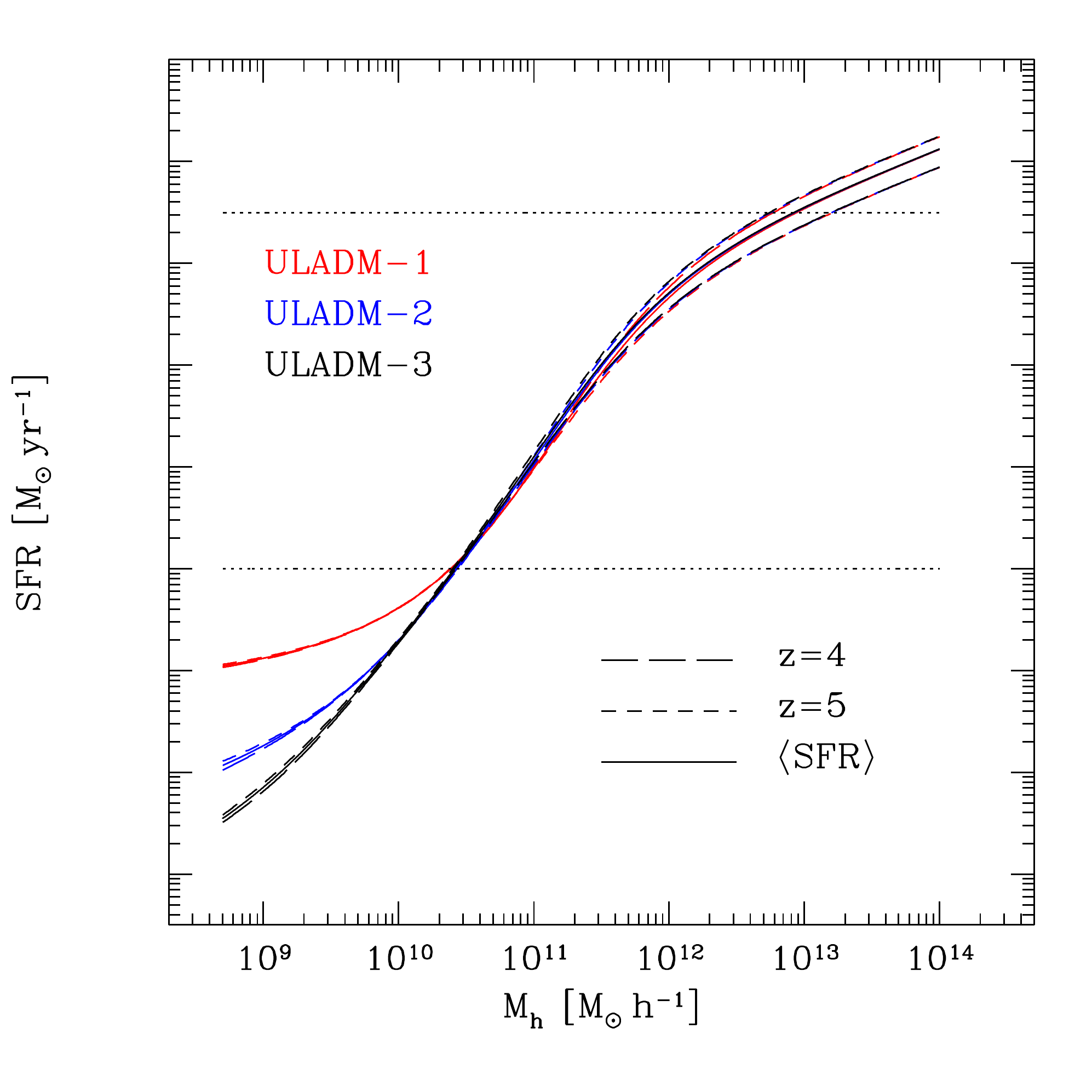}}\quad
\caption{\label{fig7} ${\rm M_h}$-SFR relation at $z=4$ (short dashed line) and $z=5$ (long dashed line) for CDM (top left panel), WDM (top right panel), LFDM (bottom left panel) and ULADM (bottom right panel). The solid line in each panel represents the redshift-averaged relation ${\rm SFR_{av}}({\rm M_h})$ at fixed halo mass between $z=4$ and $5$, which we use as a template to model the ensemble average relation at higher redshifts (see Section~\ref{LFmodel}).}
\end{figure*}

We use the analytical fits to the SFR density functions and the halo mass functions to compute the ${\rm M_h}-{\rm SFR}$ relation at $z=4$ and $5$ from halo abundance matching,  $n(>{\rm M_h})=\Phi(>{\rm SFR})$ (see e.g. \cite{Mashian2016}), for each DM model in our simulation suite. The inferred relations are shown in Fig.~\ref{fig7} for the CDM (top left panel), WDM (top right panel), LFDM (bottom left panel) and ULADM (bottom right panel). The horizontal dotted lines indicate the limiting values of SFR covered by the extinction corrected LF measurements \cite{Bouwens2015}. 

Let us focus on the CDM case. We can see that in the range covered by the LF observations the values of SFR at $z=4$ and $5$ span three orders of magnitude, yet the difference in SFR at fixed halo mass between $z=4$ and $5$ does not exceed a factor of $2$ across the entire mass range. The largest differences occur at the high and low-mass ends. At both redshifts the curve exhibit a change of slope at ${\rm M_h^*}\approx 2\cdot 10^{11}$ M$_\odot\,h^{-1}$, with a steep power law behaviour for ${\rm M_h}<{\rm M_h^*}$ and a flatter trend for ${\rm M_h}>{\rm M_h^*}$. This is consistent with the findings of \cite{Mashian2016}, where the authors have pointed out that the steep slope at low masses can be indicative of feedback mechanisms that inhibit star formation. 

However, from Fig.~\ref{fig7} we can see that such interpretation only holds for the CDM scenario. In fact, for WDM, LFDM and ULADM models we find a systematic deviation from the CDM trend at low masses. Only at large masses, ${\rm M_h}\gtrsim 10^{11}$ M$_\odot\,h^{-1}$ all models converge to the same SFR-M$_h$ relation at both redshifts. Notice that such deviations correlate with the level of suppression of the low-mass halo abundance compared to the CDM model. For instance, in the case of WDM-1 the SFR in halos of mass ${\rm M_h}\approx 10^{10}$ M$_\odot\,h^{-1}$ is up to a factor $\sim 10$ larger than CDM, while in the case of WDM-2 the SFR is a factor $\sim 3$ larger. This follows from the imposed equality of the number density distributions in halo mass and SFR. In other words, DM models with suppressed halo mass abundance at fixed low mass need a larger star formation rate to match the observed SFR densities compared to the CDM prediction. The end result is that in the alternative DM scenarios the star formation rate halo mass relation tends to flatten at the low mass end compared to CDM model. This implies that in such alternative DM scenarios, in order to reproduce LF observations, feedback mechanisms must operate differently than in CDM, being either less efficient in suppressing star formation or even promoting it depending on the specificities of DM model considered. The plots shown in Fig.~\ref{fig7} also suggest that LF independent measurements of the SFR and host halo mass in galaxies at $z=4$ and $5$ can directly test these trends and put constraints on the DM scenario. 

An important point that we would like to emphasise is the fact that the shape of the ${\rm M_h}-{\rm SFR}$ remains mostly unaltered between $z=4$ and $5$, which corresponds to a time scale of $\approx 360$ Myr. In particular, the differences between the two curves can be accounted to good approximation by an overall amplitude factor. Thus, for each DM model, we derive a template form of the ${\rm M_h}-{\rm SFR}$ relation, ${\rm SFR_{av}}({\rm M_h})$, by averaging the HAM inferred relation at $z=4$ and $5$ at fixed halo mass. We use this template function to model the ensemble average star formation rate mass relation at $z=6,7$ and $8$. This is similar in spirit to the LF model calibration by Mason, Trenti \& Treu \cite{Mason2015} who have performed HAM against LF data at $z=5$ from \cite{Bouwens2015} to calibrate a redshift independent star formation efficiency. In our case, we assume that the ensemble average star formation rate differs from the calibrated template at $z=6,7$ and $8$ by an unknown constant amplitude factor. 

This {\rm ansatz} has a twofold aspect. First, it implies that the shape of ${\rm M_h}-{\rm SFR}$ over the mass range of interest is set by physical mechanisms very early on at redshift higher than $z=8$. Secondly, since $z=5$ and $8$ differ by approximately $\approx 500$ Myr, the assumption of a constant scaling in amplitude is equivalent to assuming that there are no feedback processes (e.g. supernova explosions) that in such a time scale can significantly alter the shape of the average ${\rm M_h}-{\rm SFR}$ relation in the mass range of interest.

\subsection{Modelling of high-$z$ luminosity function}\label{LFmodel}
We model the galaxy LF at $z>5$ as in Mashian et al. \cite{Mashian2016} and assume that the probability of having a galaxy with star formation rate SFR in a halo of mass M$_h$ (i.e. the conditional SFR density) is given by a log-normal distribution. However, differently from \cite{Mashian2016} we assume the mean\footnote{We denote as $\langle{\rm SFR}({\rm M_h},z)\rangle$ the ensemble average of the SFR at fixed halo mass ${\rm M_h}$ and redshift $z$.} to be given by $\langle {\rm SFR}({\rm M_h},z)\rangle\equiv\varepsilon^z_{\rm SFR}{\rm SFR_{av}(M_h)}$ with variance $\sigma^2_{z,{\rm SFR}}$:
\begin{equation}
P({\rm SFR|{\rm M_h}})=\frac{e^{-\frac{1}{2\sigma^2_{z,{\rm SFR}}}\log^2_{10}\left[\frac{\rm SFR}{\langle{\rm SFR}({\rm M_h},z)\rangle}\right]}}{{\rm SFR}\sqrt{2\pi\sigma^2_{z,{\rm SFR}}}},\label{lognorm}
\end{equation}
where ${\rm SFR_{av}(M_h)}$ is the template relation previously computed for each DM model, $\varepsilon^z_{\rm SFR}$ is a free parameter accounting for the overall amplitude of the $\langle {\rm SFR}({\rm M_h},z)\rangle$ relation at redshift $z$ and $\sigma^z_{\rm SFR}$ is a free parameter which accounts for the intrinsic scatter around this relation. The latter parametrises the stochasticity of the processes which are responsible for the star formation. 

We compute the SFR density at a given redshift by integrating Eq.~(\ref{lognorm}) over the halo mass function of a given DM model:
\begin{equation}
\phi({\rm SFR},z)= \int \frac{dn}{d{\rm M_h}}({\rm M_h},z) P({\rm SFR |{\rm M_h}}) d{\rm M_h},
\end{equation}
then using the Kennicutt-relation we convert SFR into UV-magnitudes to derive the extinction-free LF function in terms of the SFR density\footnote{For clarity, we have
\begin{equation}
\phi({\rm M_{UV}^c})=\frac{d\Phi}{d{\rm SFR}}\frac{d {\rm SFR}}{d L_{UV}}\left\lvert\frac{d L_{UV}}{d{\rm M_{UV}^c}}\right\rvert=\frac{1}{2.5}{\rm SFR}\frac{d\Phi}{d{\rm SFR}}\ln(10),\nonumber
\end{equation}
where in the last term we have used the Kennicutt-relation and the magnitude-luminosity relation.
},
\begin{equation}
\phi({\rm M_{UV}^c})\equiv\frac{1}{2.5}\,\varphi({\rm SFR[{\rm M_{UV}^c}]},z).\label{phimodel}
\end{equation}
Finally, using Eq.~(\ref{phimuvc}) to transform dust-free UV-magnitudes into the observed ones, $\phi({\rm M_{UV}^c})\rightarrow \phi({\rm M_{UV}})$, we obtain the DM model prediction of the LF with $\varepsilon^z_{\rm SFR}$ and $\sigma^z_{\rm SFR}$ as free parameters of the model.

At this point it is worth reminding the reader of the differences between the approach described above and the evaluation of the high-z LF presented in \cite{Mashian2016} and \cite{Schive2016}. In \cite{Mashian2016}, the ensemble average $\langle {\rm SFR}({\rm M_h},z)\rangle$ is derived by averaging at fixed halo mass the ${\rm M_h}-{\rm SFR}$ relation inferred from halo abundance matching using SFR densities estimates from LF measurements at $4\le z\le 8$. Moreover, they set a redshift-independent intrinsic scatter to $\sigma_{\rm SFR}=0.5$ dex. In contrast, in \cite{Schive2016} the authors have modelled the conditional luminosity function as a log-normal distribution and assumed a parametric form of the average UV-luminosity halo mass relation (with three fitting model parameters) and a free dispersion parameter inspired by the conditional LF model of \cite{Cooray2005}.

Our approach takes advantage of both methods in that the ensemble average star formation halo mass relation is modelled in terms of a template function (rather than a parametric one) determined from LF data at $z\approx 4-5$ (rather than averaging across a larger redshift interval) with the LF prediction depending only on two free model parameters.

\section{Data Analysis}\label{data_analysis}
\subsection{Luminosity function datasets}
Over the past few years several groups have measured the galaxy luminosity function at $z\gtrsim 6$ (see e.g. \cite{Mclure2013,Schenker2013,Bouwens2015,Finkelstein2015}). These studies have been able to precisely characterise the bright-end of the LF (${\rm M_{UV}}<-15$), while only recently it has been possible to probe the faint-end thanks to the detection of faint distant objects lensed by massive clusters observed in the context of {\it Hubble} Frontier Fields program \cite{Atek2015b,Livermore2016,Bouwens2016c}. 

Bouwens et al. \cite{Bouwens2015} (B15 hereafter) have characterised the luminosity function at $4\lesssim z\lesssim 10$ using the largest galaxy sample to date from {\it HST} data. In Section~\ref{sfrav} we have used their LF measurements at $z=4$ and $5$ to calibrate our SFR-M$_h$ template. We use their LF estimates at $z=6,7$ and $8$ (see Table 5 in B15) to infer constraints on the simulated DM models. 

Previous estimate of the LF in the same magnitude range at $z=7$ and $8$ were obtained by Schenker et al. \cite{Schenker2013} and McLure et al. \cite{Mclure2013}. However, these analyses use smaller galaxy samples than those of B15. Moreover, their estimates of the total magnitude of galaxies assume them to be point sources, which as shown in the analysis presented in the Appendix G of \cite{Bouwens2015} introduces a systematic bias in the UV-magnitude determination of the most luminous and extended sources. For these reasons we do not include the LF measurements from \cite{Schenker2013,Mclure2013} in our analysis.  

Luminosity function measurements over the same redshifts and UV-magnitude intervals of B15 have been obtained
independently by Finkelstein et al. \cite{Finkelstein2015}. The galaxy samples used in the latter work largely overlap with those of B15. However, this analysis uses different selection criteria and data reduction procedures which may be responsible for the slight underabundance of bright galaxies compared to the findings of B15. 

The dust extinction model parameters used in our analysis have been calibrated to the values inferred from \cite{Bouwens2014} which uses galaxy samples (and data reduction procedures) that are common to those used in B15. Furthermore, the effect of the extinction is more important on the bright-end of the LF. Hence, for coherence we discard the LF data from Finkelstein et al. \cite{Finkelstein2015} and only use the B15 data to cover the bright-end interval of the LF. 

Measurements of the bright-end LF at $z=6$ and $7$ have also been obtained by Willott et al. \cite{Willott2013} and Ouchi et al. \cite{Ouchi2009} respectively. More recently Bowler et al. \cite{Bowler2014,Bowler2015,Bowler2017} have derived LF estimates at these redshifts in the magnitude range $-24\lesssim {\rm M_{UV}}\lesssim -22$, which in combination with the B15 data at $z=6$ and $7$ better anchor the bright-end LF slope. However, we have verified that the use of these additional datasets does not lead to further constraints on the DM model fits, since these only affect the faint-end of the LF.

LF measurements to ${\rm M_{UV}}\approx -15$ have been obtained by Atek et al. (\cite{Atek2015b}, A15 hereafter) using a sample of $227$ galaxy candidates at $z\sim 6-7$ and $25$ candidates at $z\sim 8$ detected through \textit{HST} observations of A2744, MACS 0416 and MACS 0717 clusters from the HFF program. More recently Livermore, Finkelstein \& Lotz (\cite{Livermore2016}, L16 hereafter) have obtained LF measurements to ${\rm M_{UV}}= -12.5$ at $z \sim 6$, ${\rm M_{UV}}=-14$ at $z \sim 7$ and ${\rm M_{UV}}=-15$ at $z \sim 8$ using a sample of $167$ galaxies detected through the analysis of A2744 and MACS 0416 clusters in HFF and including sources that escaped detection in A15.

As mentioned in Section~\ref{intro} these observations have opened the way to a new alternative approach to explore the population of faint sources in the far distant universe. However, differently from deep survey searches, these LF estimates may be affected by a number of systematic errors that need to be carefully evaluated. As pointed out by Bouwens et al. (\cite{Bouwens2016c}, hereafter B16), lens mass model uncertainties affect the evaluation of the magnification of the faintest sources and introduce a systematic bias in the characterisation of the LF. In B16, the authors have introduced a methodology to incorporate this effect in the determination of the LF. Using a sample of $160$ galaxies detected through observations of Abell 2744, MACS 0416, MACS 0717 and MAC 1149 they inferred the LF at $z=6$. 

\begin{table}[t]
\centering
\begin{tabular}{|c|c|c|c|}
\hline\hline
Dataset & $\alpha_{\rm LF}^{z=6}$ & $\alpha_{\rm LF}^{z=7}$& $\alpha_{\rm LF}^{z=8}$ \\
\hline
B15 \cite{Bouwens2015}&   $-1.90\pm 0.10$  &  $-2.06\pm 0.13$ & $-2.02\pm 0.23$\\
\hline
A15 \cite{Atek2015b}&   - & $-1.98\pm 0.12$ & $-2.23\pm 0.37$\\
\hline
L16 \cite{Livermore2016}& $-2.18\pm 0.13$ & $-2.05\pm 0.24$ & $-1.99\pm 0.26$ \\
\hline
B16 \cite{Bouwens2016c} & $-1.84\pm 0.06$ & - & - \\
\hline\hline
\end{tabular}
\caption{Marginalised mean and $1\sigma$ error on $\alpha_{\rm LF}$ inferred from the likelihood analysis of the LF datasets at $z=6,7$ and $8$.\label{tab_alphalf}.}
\end{table}
\begin{figure}
\includegraphics[scale=0.4]{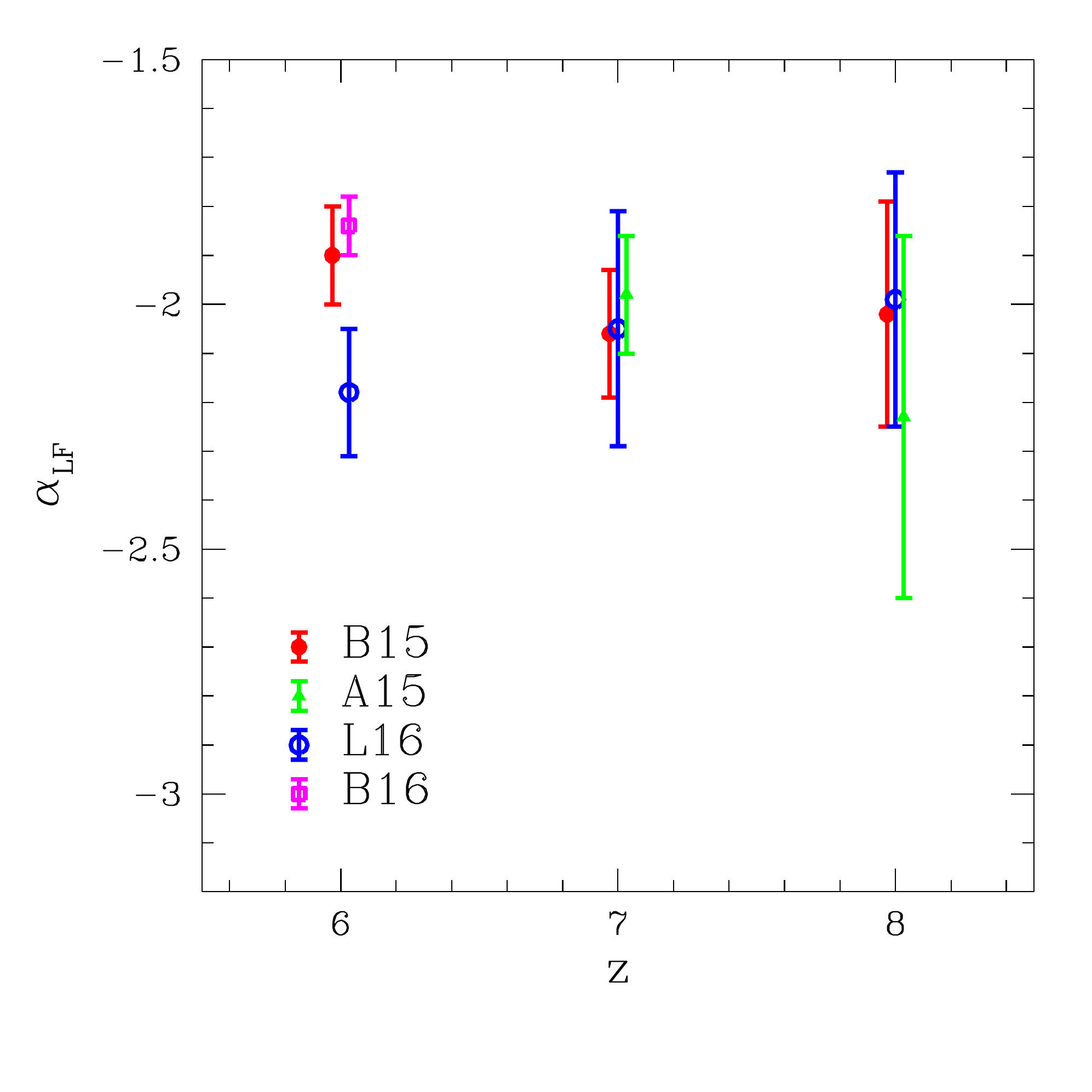}
\caption{Marginalised mean and $1\sigma$ error on $\alpha_{\rm LF}$ as inferred from the likelihood analysis of the Schechter-fit to the LF measurements from B15 (red solid circles), A15 (green solid triangles), L16 (blue empty circles), B16 (magenta empty squares) at $z=6,7$ and $8$. The points have been slightly displaced in redshift to facilitate readability of the figure.}\label{fig_alpha_slope}
\end{figure}

As a consistency check, we fit the various LF data with the Schechter-function:
\begin{equation}\label{schechter}
\phi({\rm M_{UV}})=\phi_{\rm LF}^*\frac{\ln{(10)}}{2.5}\frac{e^{-10^{-0.4({\rm M_{UV}-M_{UV}^*})}}}{10^{0.4({\rm M_{UV}-M_{UV}^*})(\alpha_{\rm LF}+1)}},
\end{equation}
where $\phi_{\rm LF}^*$ is a normalisation parameter, $M_{UV}^*$ an exponential cut-off scale and $\alpha_{\rm LF}$ the faint-end slope, and run a Markov Chain Monte Carlo (MCMC) likelihood analysis to infer constraints on the Schechter parameters for the B15, A15, L16 and B16 datasets . We refer the reader to Section~\ref{stat} for a general description of the statistical approach adopted here. Since constraints on the different DM models are sensitive to the faint-end slope, we do not compare constraints on $\phi_{\rm LF}^*$ and ${\rm M_{UV}^*}$ across the datasets and limit our consistency test to $\alpha_{\rm LF}$. 

In Table~\ref{tab_alphalf} we quote the marginalised mean and $1\sigma$ error on $\alpha_{\rm LF}$. Notice that the mean values given in A15, L16, and B16 are slightly different from those quoted in Table~\ref{tab_alphalf}. More importantly our estimated errors on $\alpha_{\rm LF}$ are larger. This is because differently from the original analyses we do not combine the datasets with additional LF measurements covering the bright-end of the luminosity function.  

In Fig.~\ref{fig_alpha_slope} we plot the values quoted in Table~\ref{tab_alphalf} for B15 (red solid circles), A15 (green solid triangles), L16 (blue empty circles), B16 (magenta empty squares). We can see that the bounds on $\alpha_{\rm LF}$ at $z=6$ from L16 do not overlap with those obtained using B15 measurements. In contrast, the LF measurements from B16, which have been obtained by accounting for the lens mass model uncertainties, give bounds that are consistent with those obtained from the fit to B15. Similarly, at $z=7$ and $8$ the values of $\alpha_{\rm LF}$ from B15, A15 and L16 are compatible with each other to within $1\sigma$. 

In the light of these results, we consider a compilation of LF data consisting of $26$ measurements (B15+B16) at $z=6$, $31$ measurements (B15+A15+L16) at $z=7$ and $22$ measurements  (B15+A15+L16) at $z=8$ spanning the UV-magnitude interval $-23\lesssim {\rm M_{UV}} \lesssim -13$, while in a separate analysis we evaluate the impact of the L16 measurements at $z=6$ on the DM models.

\begin{table*}[ht]
\centering
\begin{tabular}{|c|c|c|c||c|c|c||c|c|c||c|}
\hline\hline
Model & $\log_{10}\varepsilon_{\rm SFR}^{z=6}$ & $\log_{10}\sigma_{\rm SFR}^{z=6}$ & $\chi^2_{z=6}$ & $\log_{10}\varepsilon_{\rm SFR}^{z=7}$ & $\log_{10}\sigma_{\rm SFR}^{z=7}$ & $\chi^2_{z=7}$ &  $\log_{10}\varepsilon_{\rm SFR}^{z=8}$ & $\log_{10}\sigma_{\rm SFR}^{z=8}$ & $\chi^2_{z=8}$  & $\chi^2_{\rm tot}$\\
\hline
{\rm \textbf{CDM}} & $-0.80$ & $-0.23$ & 21.5  &$-0.52$ & $-0.23$ & 27.6 & $-0.18$ & $-0.39$ &  15.8 & {\bf 64.9} \\
\hline
{\rm WDM-1} & $-0.78$ & $-0.26$ & 87.7 & $-0.60$ & $-0.25$ & 57.4 & $-0.58$ & $-0.24$ & 23.8 & 168.9 \\
{\rm WDM-2} & $-0.79$ & $-0.25$ & 22.4 & $-0.53$& $-0.24$ & 31.6 & $-0.12$ & $-0.47$ & 17.5 & 71.5 \\
{\rm {\bf WDM-3}} & $-0.83$ & $-0.22$ & 20.5 & $-0.54$ & $-0.23$ & 28.1 & $-0.23$ & $-0.37$ & 15.7 & {\bf 64.3} \\
{\rm WDM-4} & $-0.90$ & $-0.20$ & 21.7 & $-0.60$ & $-0.21$ & 27.8 & $-0.28$ & $-0.35$ & 15.9 & 65.4 \\
{\rm WDM-5} & $-0.85$ & $-0.21$ & 22.0 & $-0.60$ & $-0.21$ & 27.1 & $-0.26$ & $-0.36$ & 15.8 & 64.9 \\
\hline
{\rm LFDM-1} & $-0.92$ & $-0.17$ & 37.2 & $-0.73$ & $-0.14$ & 45.0 & $-0.29$ & $-0.50$ & 16.6 & 98.8 \\
{\rm {\bf LFDM-2}} & $-0.83$ & $-0.23$ & 20.1 & $-0.53$ & $-0.23$ & 28.3 & $-0.22$ & $-0.38$ & 15.6 & {\bf 64.0} \\
{\rm LFDM-3} & $-0.85$ & $-0.22$ & 21.7 & $-0.73$ & $-0.20$ & 30.5 & $-0.49$ & $-0.29$ & 16.2 & 68.4 \\
\hline
{\rm ULADM-1} & $-0.91$& $-0.24$ &21.3 & $-0.69$ & $-0.24$ & 33.6 & $-0.48$ & $-0.36$ & 14.9 & 69.8 \\
{\rm ULADM-2} & $-0.89$& $-0.20$ &21.5 & $-0.78$ & $-0.19$ & 31.4 & $-0.59$ & $-0.26$ & 16.5 & 69.4 \\
{\rm \textbf{ULADM-3}} & $-0.81$& $-0.23$ &21.9 & $-0.60$ & $-0.21$ & 27.3 & $-0.29$ & $-0.34$ & 15.9 & {\bf 65.1} \\
\hline\hline
\end{tabular}
\caption{\label{tab4} Best-fit values of $\log_{10}\varepsilon_{\rm SFR}$ and $\log_{10}\sigma_{\rm SFR}$ at $z=6,7$ and $8$ for the various DM models with the corresponding values of the $\chi^2$. In bold letters are the models with lowest $\chi^2$ within each DM scenario.}
\end{table*}

\subsection{Likelihood Evaluation}\label{stat}
We perform a Markov Chain Monte Carlo (MCMC) likelihood data analysis to derive constraints on the parameters characterising the theoretical LF model $\phi_{\rm th}$. To this purpose we evaluate the following $\chi^2$:
\begin{equation}\label{chi2}
\chi^2= \sum_i \left[\frac{\log_{10}(\phi_{\rm obs}^i)-\log_{10}(\phi_{\rm th}^i)}{\sigma^i_{\log_{10}(\phi)}}\right]^2,
\end{equation}
where $\phi_{{\rm obs}}^i$ are the LF measured values. Since no information is available concerning possible correlations among different UV-magnitude bins, for simplicity we assume all LF measurements to be statistically independent.

We generate the random chains using a Metropolis-Hastings algorithm. We evaluate the rejection rate every 100 steps and adjust the width of the parameters dynamically. We set uniform priors for the LF model parameters. 

For the Schechter-fitting function analysis described above, we have computed the $\chi^2$ with $\phi_{\rm th}$ given by Eq.~(\ref{schechter}) and sampled a 3-dimensional parameter space with $(\log_{10}\phi^*_{\rm LF}, \alpha_{\rm LF}, \log_{10}{\rm M_{UV}^*})$ uniformly varying in the range $[-4,-2]$, $[-3,-1]$ and $[-30,-15]$ respectively. For the DM model analysis, we have computed the $\chi^2$ with $\phi_{\rm th}$ given by Eq.~(\ref{phimodel}) (after having converted the dust-free UV-magnitude into observed ones) and sampled the 2-dimensional parameter space with $(\varepsilon_{\rm SFR},\sigma_{\rm SFR})$ uniformly varying in the range $[-3,1]$. For each model we run 3 independent chains of $10^6$ samples and check their convergence using the Gelman-Rubin test \cite{Gelman1992}.

\section{Results}\label{result}
Here we present the results of the likelihood data analysis. We first focus on the goodness-of-fit of the different DM models and derive constraints on a given DM scenario using deviance statistics. Then we discuss the constraints on the parameters of the SFR-M$_h$ relation for the best-fit DM models.

\subsection{DM models goodness-of-fit}
\begin{figure*}[t]
\centering
\subfigure{\includegraphics[scale=0.4]{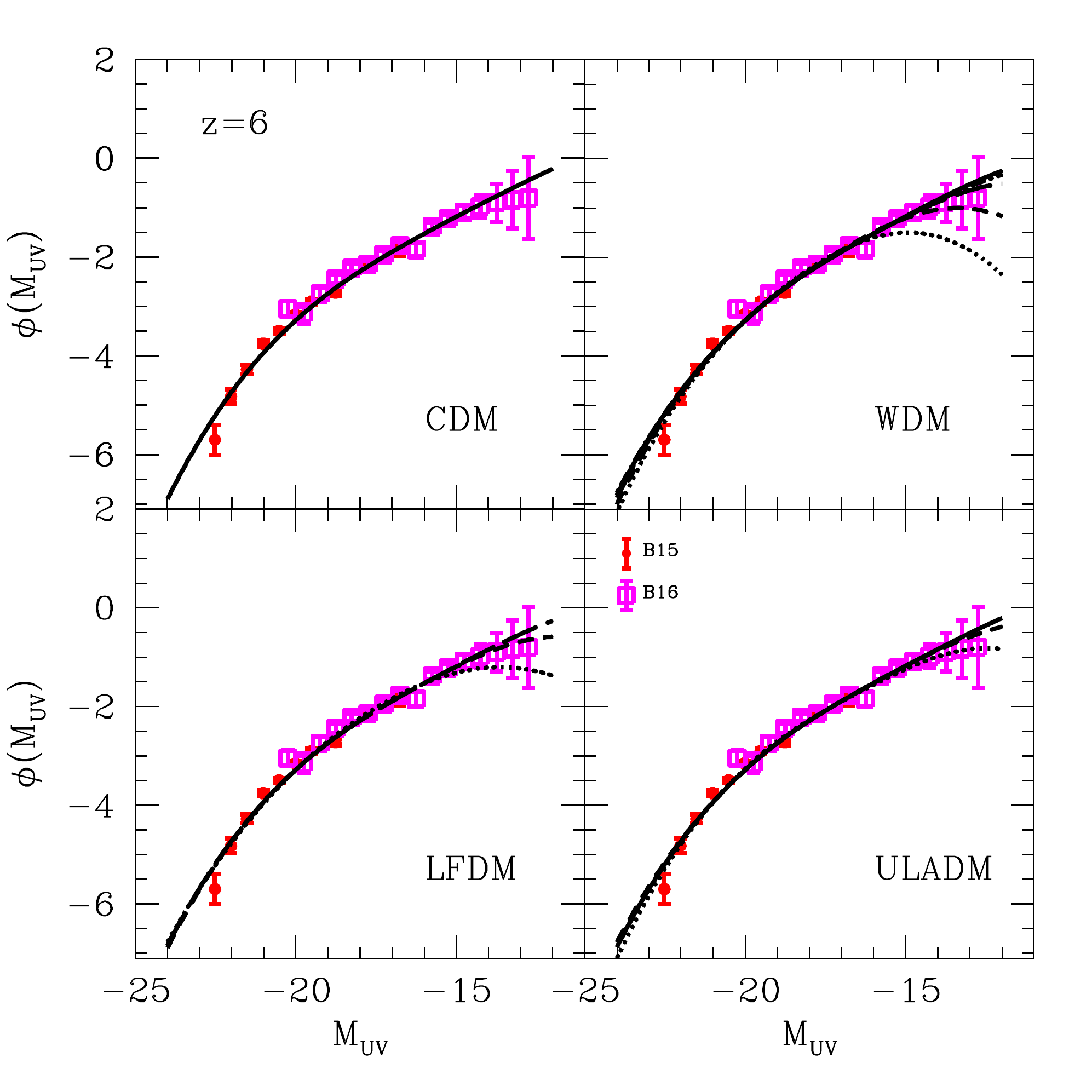}}\quad
\subfigure{\includegraphics[scale=0.4]{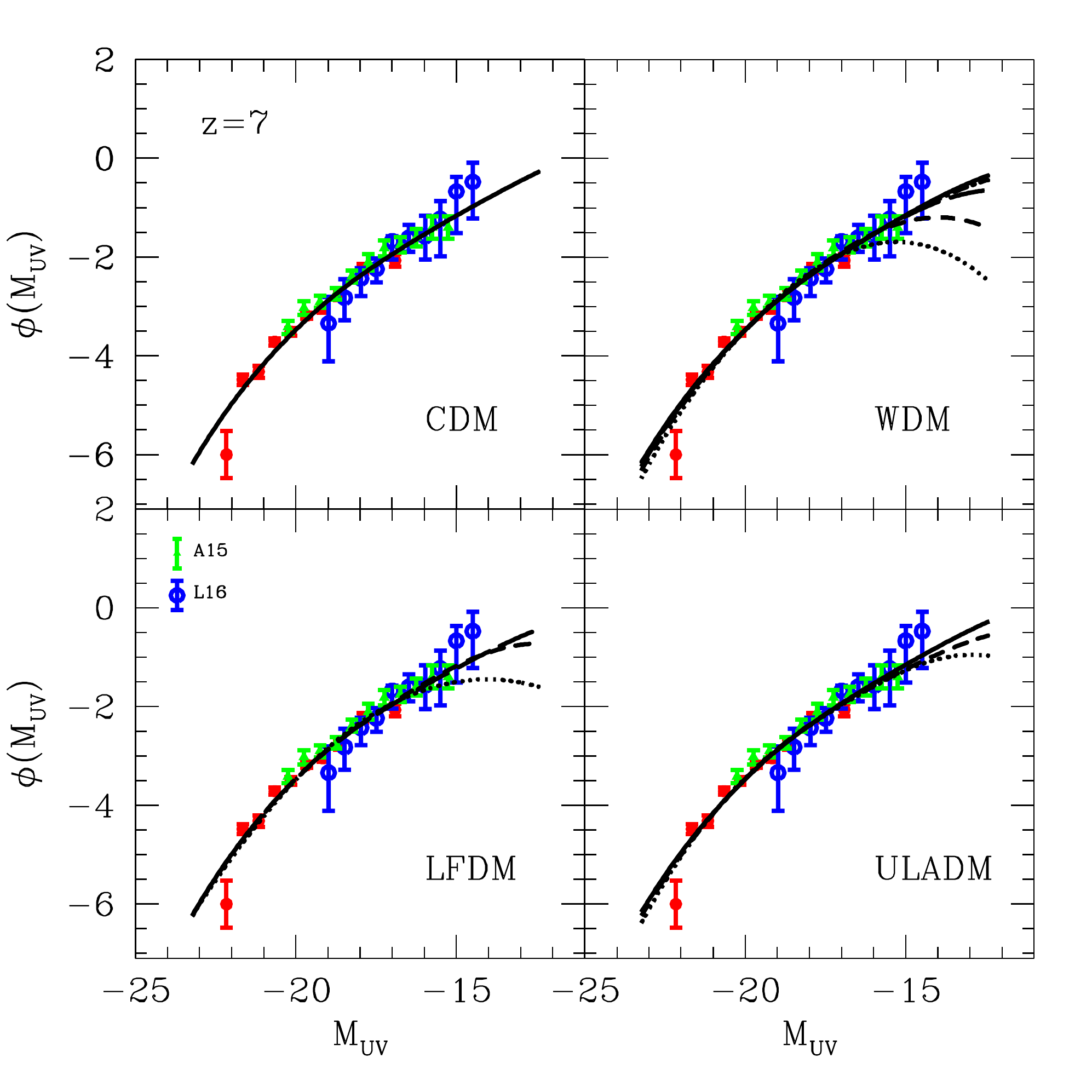}}\quad
\subfigure{\includegraphics[scale=0.4]{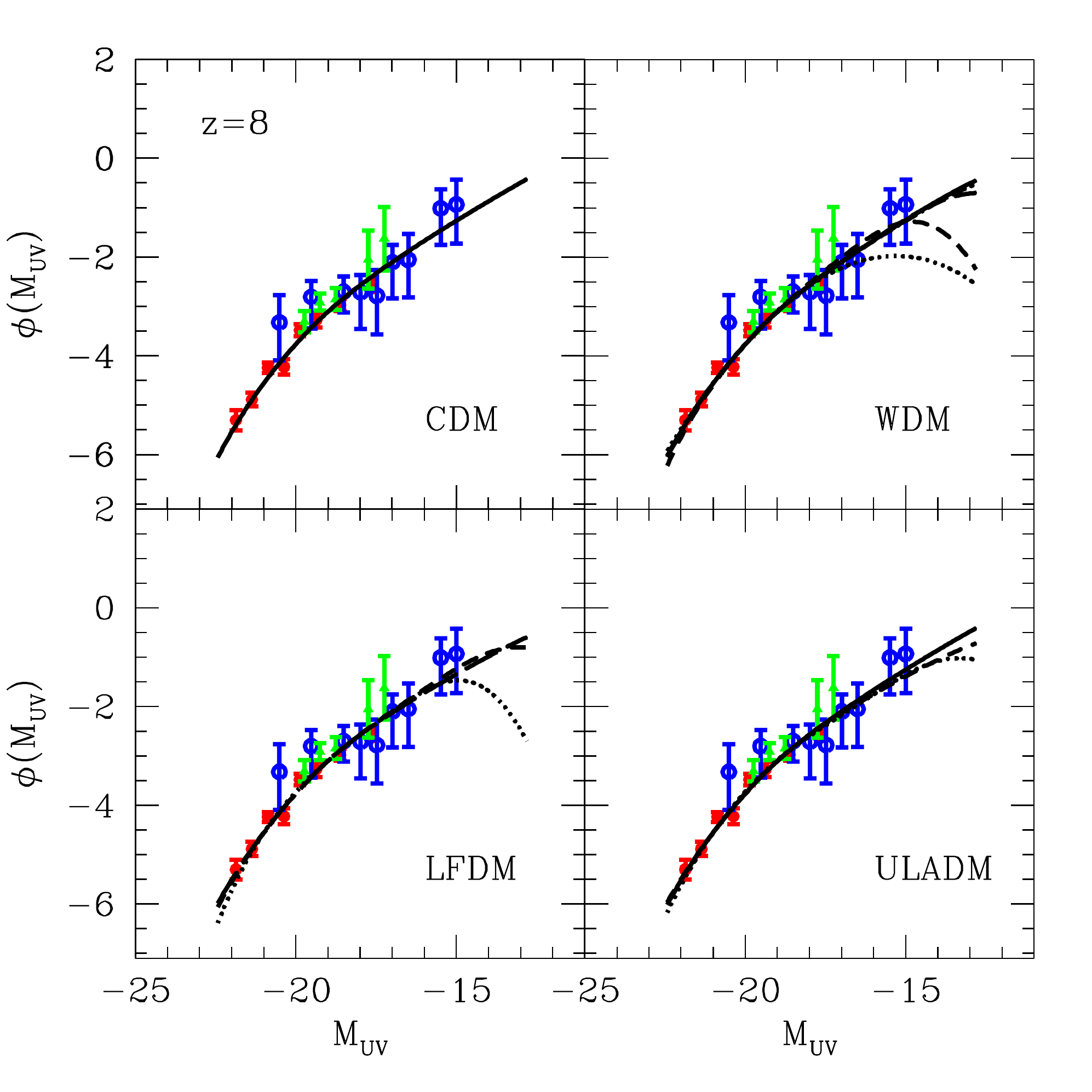}}\quad
\caption{\label{fig9} Best-fit luminosity function for the various DM models against data at $z=6$ (top left panel), $z=7$ (top right panel) and $z=8$ (bottom panel) from B15 (red filled circles), B16 (magenta empty squares), A15 (green filled triangles) and L16 (blue empty circles). Within each panel, CDM is shown in the top left; WDM models are shown in the top right (lines from bottom to top correspond to WDM-1, WDM-2, WDM-3, WDM-4 and WDM-5 respectively), LFDM models are shown in the bottom left (LFDM-1, LFDM-2 and LFDM-3 lines from bottom to top) and ULADM in the bottom right (ULADM-1, ULADM-2 and ULADM-3 lines from bottom to top).}
\end{figure*}

In Table~\ref{tab4} we quote the best-fit LF model parameters and the corresponding $\chi^2$-values at $z=6,7$ and $8$ for each of the simulated DM models, while in Fig.~{\ref{fig9}} we plot the corresponding LF against the data. 

We find that the CDM model provides a very good fit to the LF data with total reduced chi-square $\chi_{\rm tot}^2 \approx 1$. Among the alternative DM models, we find WDM-3, LFDM-2 and ULADM-3 to be those with the lowest $\chi_{\rm tot}^2$. These are comparable or even slightly lower than that of the CDM model with differences $\Delta\chi^2_{\rm tot}\lesssim 1$. Thus, given current LF measurements, these models are statistically indistinguishable from the standard CDM scenario. On the other hand, we can see that $\chi^2_{\rm tot}$ varies from one model to another within each DM scenario.

The deviance statistics indicate that the best-fit WDM-1 model is excluded at more than $5\sigma$ compared to the best-fit WDM-3 model with $\Delta\chi^2_{\rm tot}\sim 105$, WDM-2 is excluded at more than $2\sigma$ with $\Delta\chi^2_{\rm tot}\sim 7$ ($>95\%$ probability), while WDM-4 and WDM-5 are statistically compatible within $1\sigma$ ($<68\%$ probability). Thus, we infer a bound on thermal relic particle mass $m_{\rm WDM}\gtrsim 1.5$ keV at $2\sigma$. Similarly, in the LFDM case we find that LFDM-1 is excluded at more than $5\sigma$ with respect to LFDM-2 with $\Delta\chi^2_{\rm tot}\sim 35$, while LFDM-3 lies within $2\sigma$ with $\Delta\chi^2_{\rm tot}\sim 4$. This suggests that for LFDM models, the phase transition redshift $z_t\gtrsim 8\cdot 10^5$ at $2\sigma$. The best-fit models of ULADM-1 and ULADM-2 are within $2\sigma$ of ULADM-3 with differences $\Delta\chi^2_{\rm tot}\sim 4$. Thus, ultra-light axion models with $m_a\gtrsim 1.6\cdot 10^{-22}$ are compatible with the LF data well within $2\sigma$ of the deviance statistics.

It is worth noticing that the LF data at $z=6$ shows a slight preference (though not statistically significant) for DM models with a flattening of the faint-end slope. This is consistent with the results of B16 where the authors pointed out that the faint-end LF measurements at $z=6$ permit a turnover (within $1\sigma$) at $-15\lesssim {\rm M_{UV}\lesssim -14}$. At higher redshifts LF data do not cover such faint magnitude interval and have much larger statistical uncertainties, thus the presence of a turnover or a flattening in the LF remains largely uncertain. In the context of the standard CDM scenario, the flattening of LF at the faint-end and the presence of a turnover can be the signatures of physical processes affecting star formation in low-mass galaxies as found in numerical studies based on hydrodynamical simulations \cite{Jaacks2013,Oshea2015,Wise2014} and semi-analytic models of galaxy formation \cite{Liu2016}. From Fig.~\ref{fig9} we can clearly see that such a feature is also a distinct prediction of DM models alternative to the CDM paradigm.

\begin{figure}[ht]
\includegraphics[scale=0.43]{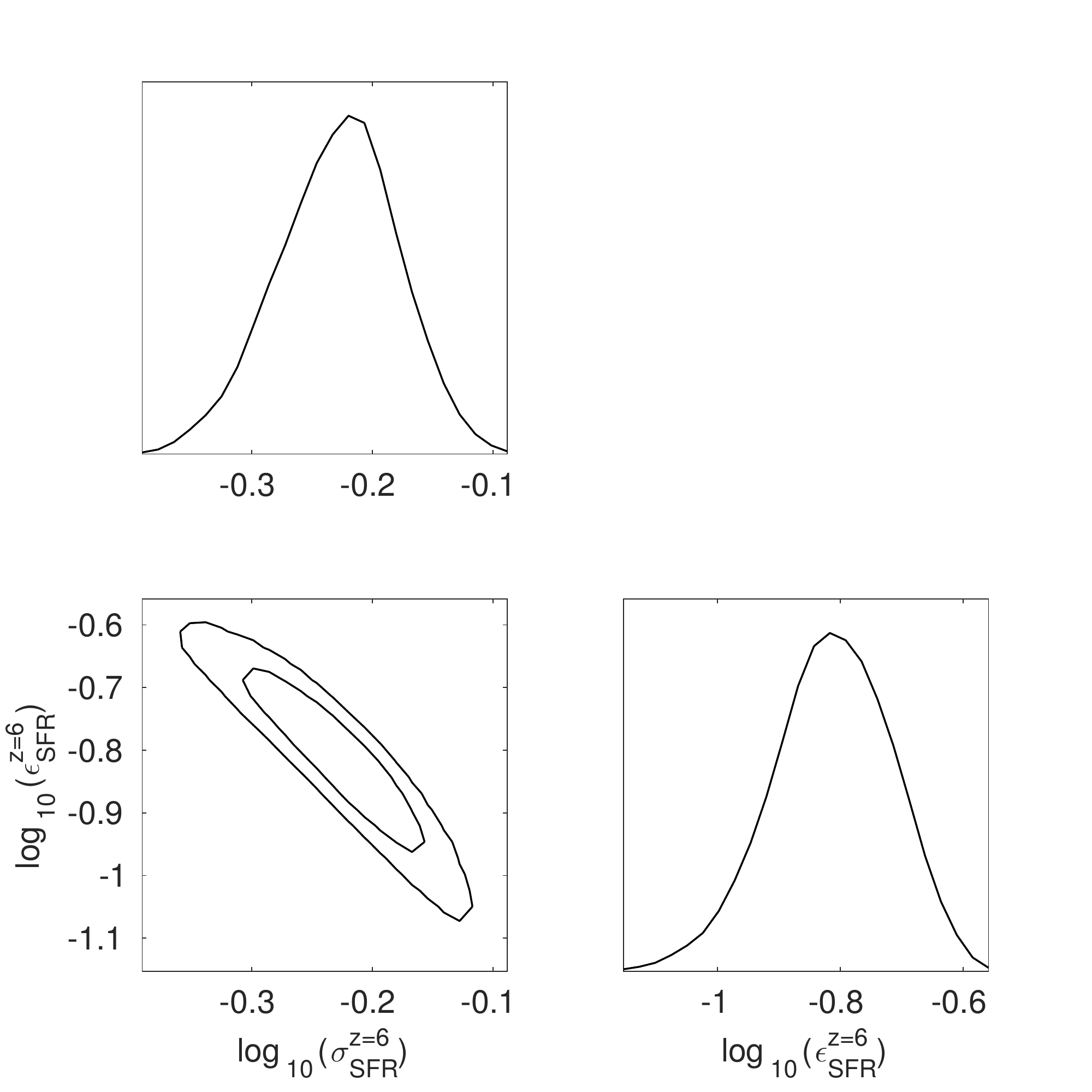}
\caption{Triangle plot of the one and two dimensional distributions of $\log_{10}\varepsilon_{\rm SFR}$ and $\log_{10}\sigma_{\rm SFR}$ for the CDM model at $z=6$. We find similar results using the data at higher redshifts and for the other DM models.}\label{triangle}
\end{figure}

\subsection{Constraints on SFR-M$_h$ relation}
We compute the average and $1\sigma$ marginalised errors on $\log_{10}\varepsilon_{\rm SFR}$ and $\log_{10}\sigma_{\rm SFR}$ from the MCMC chains. The values for the various DM models at $z=6,7$ and $8$ are quoted in Table~\ref{tab3}. We find the parameter posteriors to be well approximated by a Gaussian distribution. This can be inferred from the fact that the best-fit LF parameter values given in Table~\ref{tab4} coincide to good approximation with average ones. As an example, this can also be seen in Fig.~\ref{triangle} where we show the triangle plot of the one and two-dimensional distributions of $\log_{10}\varepsilon_{\rm SFR}$ and $\log_{10}\sigma_{\rm SFR}$ for the CDM model at $z=6$. Similar results hold for all models and at all redshifts.

\begin{figure*}[ht]
\centering
\subfigure{\includegraphics[scale=0.4]{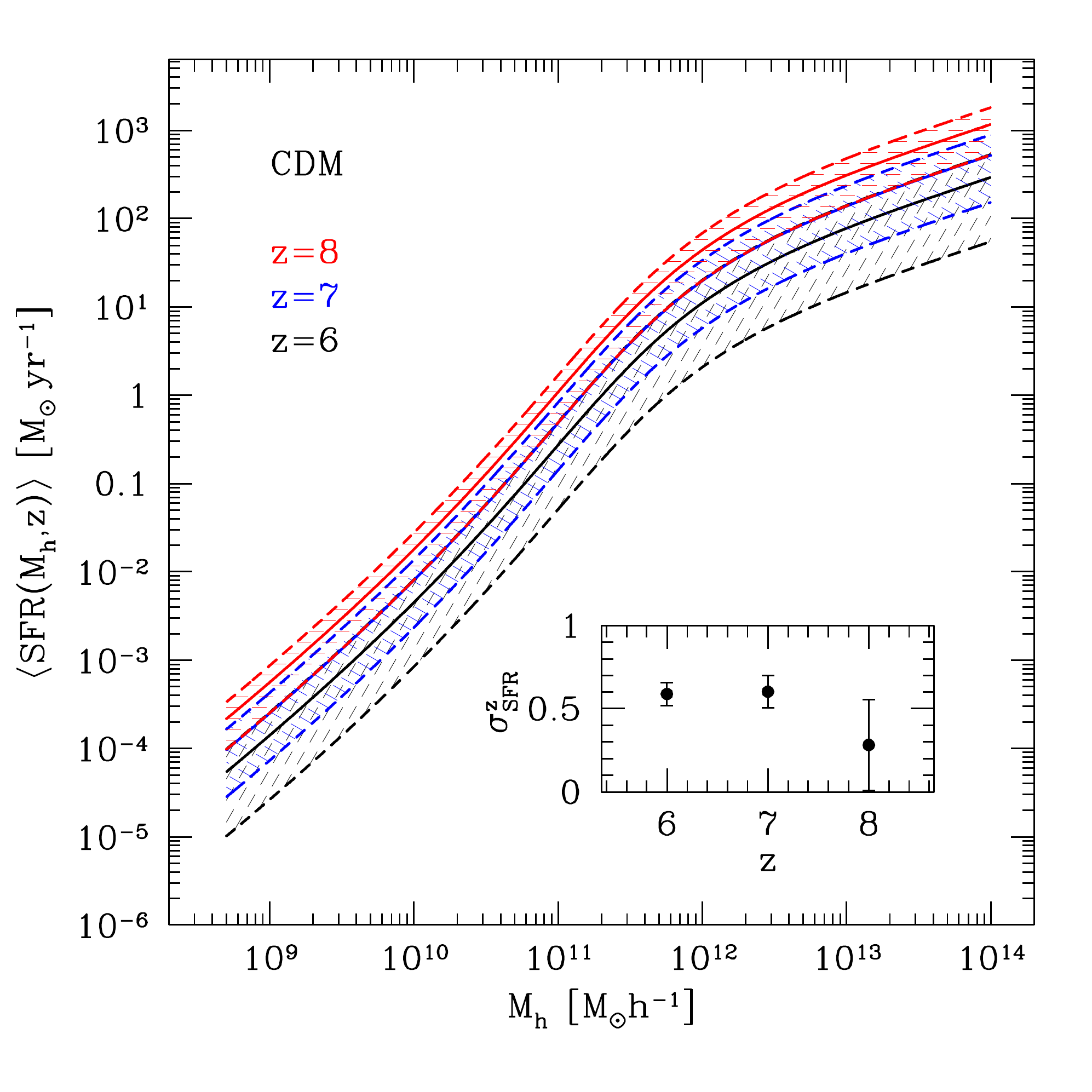}}\quad
\subfigure{\includegraphics[scale=0.4]{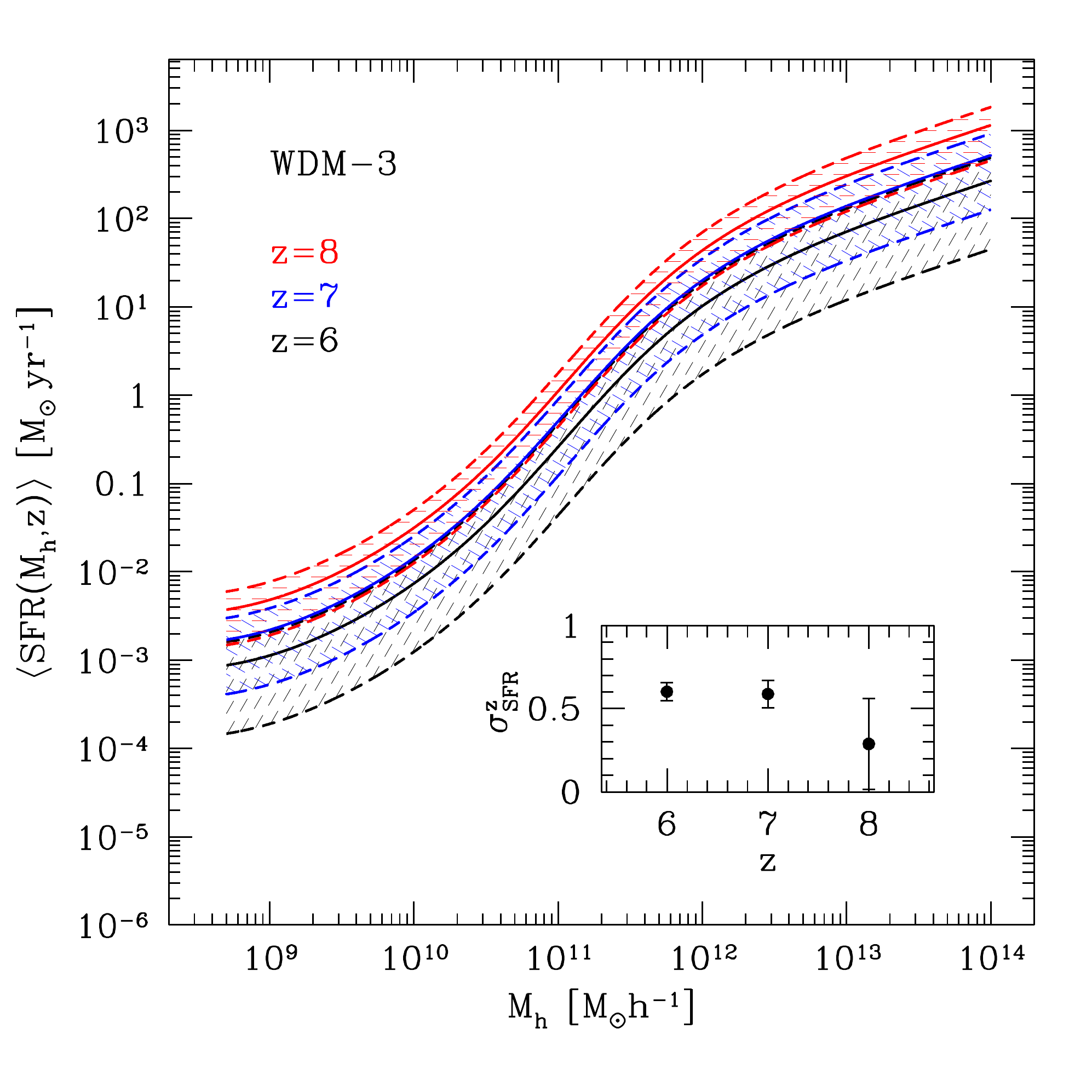}}\quad
\subfigure{\includegraphics[scale=0.4]{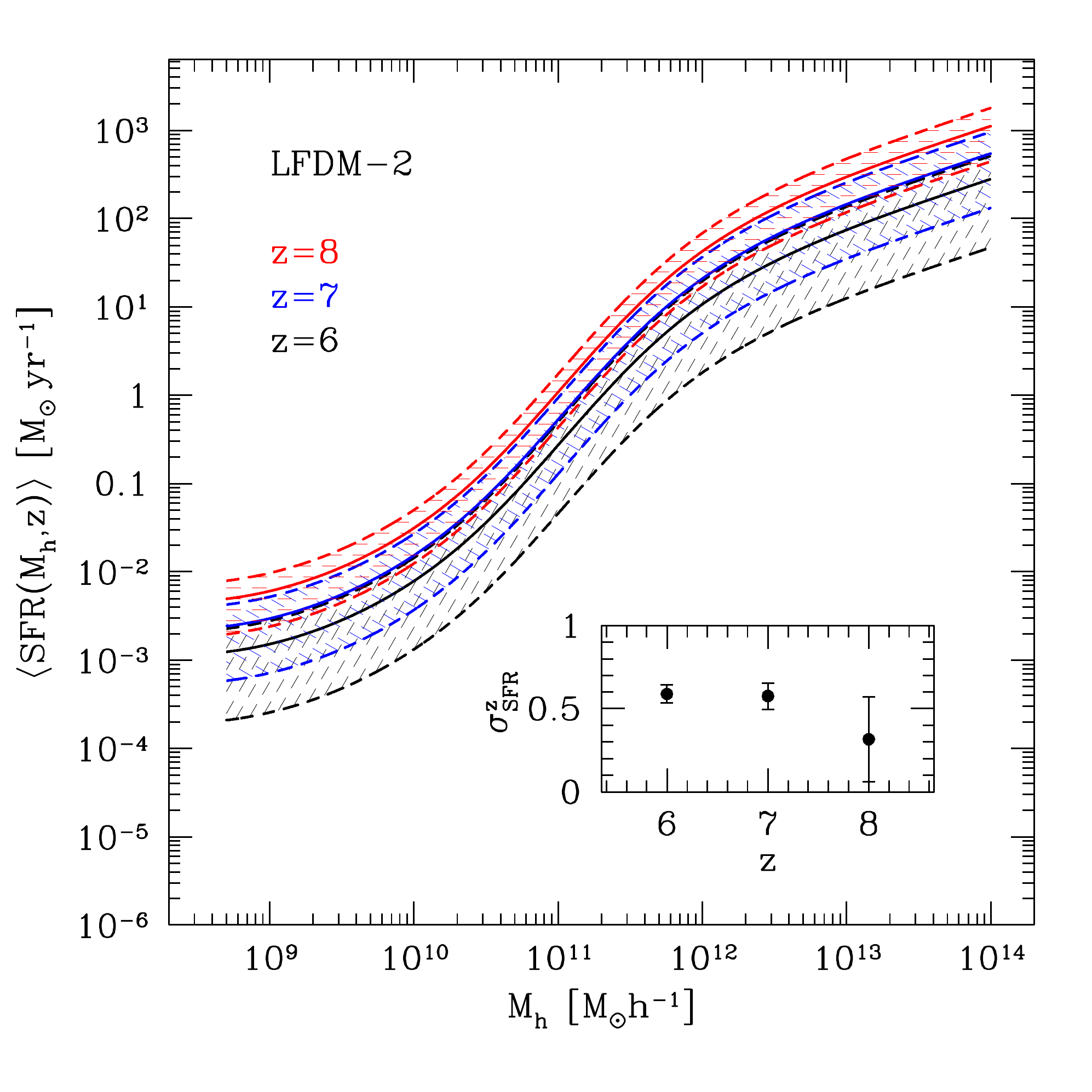}}\quad
\subfigure{\includegraphics[scale=0.4]{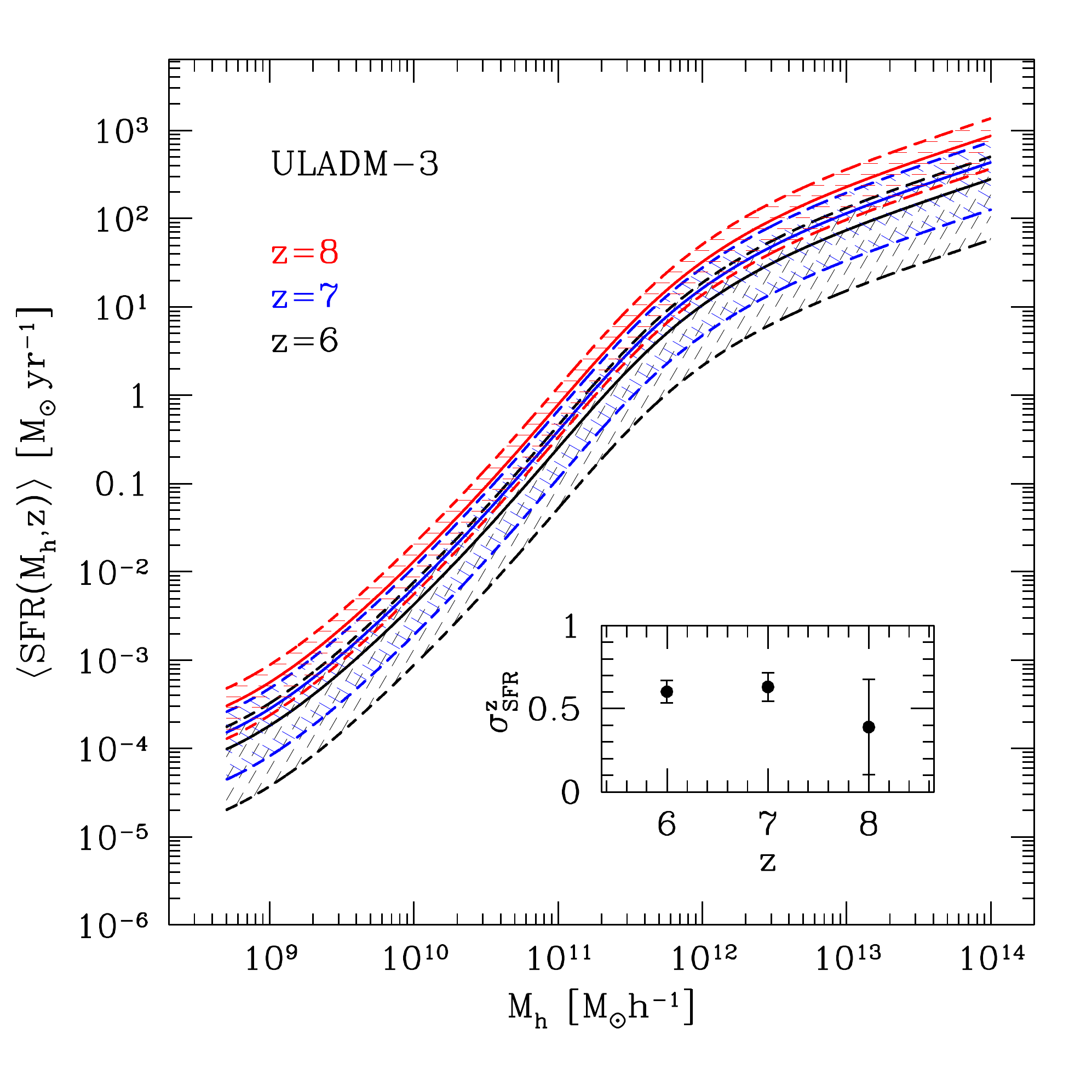}}\quad
\caption{\label{fig11} Average SFR vs halo mass, $\langle{\rm SFR(M_h),z}\rangle$ at $z=6$ (black lines), $z=7$ (blue lines) and $z=8$ (red lines) for CDM (top left panel), WDM-3 (top right panel), LFDM-2 (bottom left panel) and ULADM-3 (bottom right panel). The solid lines corresponds to the marginal mean value of $\varepsilon_{\rm SFR}^z$ while the dashed lines to the $68\%$ statistical uncertainties quoted in Table~\ref{tab3} with the hatched area in between covering the $\pm 1\sigma$ errors. In each panel we also plot the marginal mean and $1\sigma$ error on $\sigma_{\rm SFR}$ at $z=6,7$ and $8$.}
\end{figure*}

From the values quoted in Table~\ref{tab3} we can see that the values of $\varepsilon^z_{\rm SFR}$ and $\sigma^z_{\rm SFR}$ at a fixed redshift do not vary significantly among the different DM models. On the other hand, we find evidence of a systematic increase of  $\varepsilon^z_{\rm SFR}$ from $z=6$ to $z=8$, while the scatter remains constant. We can see this more clearly in Fig.~\ref{fig11} where we plot $\langle{\rm SFR(M_h),z}\rangle$ at $z=6$ (black lines), $z=7$ (blue lines) and $z=8$ (red lines) for CDM (top left panel), WDM-3 (top right panel), LFDM-2 (bottom left panel) and ULADM-3 (bottom right panel). The solid lines represent the ensemble average star formation halo mass relation with amplitude factor $\varepsilon^z_{\rm SFR}$ given by the marginal mean value quoted in Table~\ref{tab3}. The dashed lines indicate the $\pm 1\sigma$ statistical error. In each panel we also plot the marginal mean of the intrinsic scatter $\sigma_{\rm SFR}$ and the related uncertainty. 
\begin{table*}[ht]
\centering
\begin{tabular}{|c|c|c|c|c|c|c|}
\hline\hline
Model & $\log_{10}\varepsilon_{\rm SFR}^{z=6}$ & $\log_{10}\sigma_{\rm SFR}^{z=6}$ & $\log_{10}\varepsilon_{\rm SFR}^{z=7}$ & $\log_{10}\sigma_{\rm SFR}^{z=7}$ &  $\log_{10}\varepsilon_{\rm SFR}^{z=8}$ & $\log_{10}\sigma_{\rm SFR}^{z=8}$ \\
\hline
{\rm CDM} & $-0.81\pm 0.09$ & $-0.23\pm 0.05$ & $-0.56\pm 0.15$ & $-0.22\pm 0.07$ & $-0.21\pm 0.26$ & $-0.55\pm 0.42$  \\
\hline
{\rm WDM-1} & $-0.78\pm 0.04$ & $-0.26\pm 0.02$ & $-0.60\pm 0.07$ & $-0.25\pm 0.03$ & $-0.58\pm 0.14$ & $-0.24\pm 0.06$ \\
{\rm WDM-2} & $-0.79\pm 0.06$ & $-0.24\pm 0.03$ & $-0.53\pm 0.11$ & $-0.24\pm 0.06$ & $-0.14\pm 0.27$ & $-0.70\pm 0.48$ \\
{\rm WDM-3} & $-0.85\pm 0.08$ & $-0.22\pm 0.04$ & $-0.56\pm 0.12$& $-0.23\pm 0.06$ & $-0.22\pm 0.22$ & $-0.54\pm 0.41$ \\
{\rm WDM-4} & $-0.92\pm 0.09$ & $-0.20\pm 0.04$ & $-0.64\pm 0.14$ & $-0.21\pm 0.06$ & $-0.29\pm 0.23$ & $-0.44\pm 0.19$ \\
{\rm WDM-5} & $-0.86\pm 0.09$ & $-0.21\pm 0.04$ & $-0.62\pm 0.14$ & $-0.21\pm 0.06$ & $-0.27\pm 0.24$ & $-0.49\pm 0.37$ \\
\hline
{\rm LFDM-1} & $-0.93\pm 0.06$ & $-0.17\pm 0.02$ & $-0.74\pm 0.09$ & $-0.14\pm 0.03$ & $-0.29\pm 0.15$ & $-0.63\pm 0.35$ \\
{\rm LFDM-2} & $-0.83\pm 0.08$ & $-0.23\pm 0.04$ & $-0.54\pm 0.12$ & $-0.24\pm 0.06$ & $-0.23\pm 0.22$ & $-0.50\pm 0.35$  \\
{\rm LFDM-3} & $-0.86\pm 0.09$ & $-0.22\pm 0.04$ & $-0.75\pm 0.13$ & $-0.20\pm 0.05$ & $-0.51\pm 0.19$ & $-0.30\pm 0.14$ \\
\hline
{\rm ULADM-1} & $-0.90\pm 0.06$& $-0.24\pm 0.03$ & $-0.71\pm 0.10$ & $-0.24\pm 0.05$ & $-0.51\pm 0.14$ & $-0.36\pm 0.10$ \\
{\rm ULADM-2} & $-0.91\pm 0.09$& $-0.20\pm 0.04$ & $-0.80\pm 0.11$ & $-0.19\pm 0.04$ & $-0.61\pm 0.17$ & $-0.27\pm 0.09$ \\
{\rm ULADM-3} & $-0.83\pm 0.10$& $-0.22\pm0.05$ & $-0.64\pm 0.15$ & $-0.20\pm 0.06$ & $-0.34\pm 0.24$ & $-0.41\pm 0.32$ \\
\hline\hline
\end{tabular}
\caption{\label{tab3} Average and $1\sigma$ marginalised errors on $\log_{10}\varepsilon_{\rm SFR}$ and $\log_{10}\sigma_{\rm SFR}$ at $z=6,7$ and $8$ for the different DM models.}
\end{table*}

For DM models with lowest $\chi^2_{\rm tot}$, we find the intrinsic scatter $\sigma_{\rm SFR}\sim 0.6$ dex which is consistent with the value assumed in \cite{Mashian2016}. 

We may notice that all models exhibit the same power law trend at high masses, ${\rm M_h}\gtrsim 10^{12}$ M$_\odot\,h^{-1}$, while differences occur at lower masses. In particular, we notice a broken power law behaviour at the low-mass end for the WDM-5, LFDM-3 and ULADM-3 models, which have higher SFR at fixed halo mass than the CDM case. This is consistent with the conclusion of the study on early galaxy formation in the WDM scenario presented in \cite{Dayal2015} and based on semi-analytical models (see also \cite{Kang2013}). As already stressed in Section~\ref{sfrav}, the differences at low masses imply that in non-standard DM models the baryonic processes that regulate star formation during early galaxy formation must operate differently than in CDM. More precisely, the curves shown in Fig.~\ref{fig11} represent constraints that simulations including baryonic physics in such alternative DM models need to reproduce to be compatible with LF observations. This will be worth investigating in the future using DM+hydro simulations.

\section{Implications of Livermore et al. \cite{Livermore2016} data at $z=6$}
\begin{figure}[ht]
\centering
\includegraphics[scale=0.43]{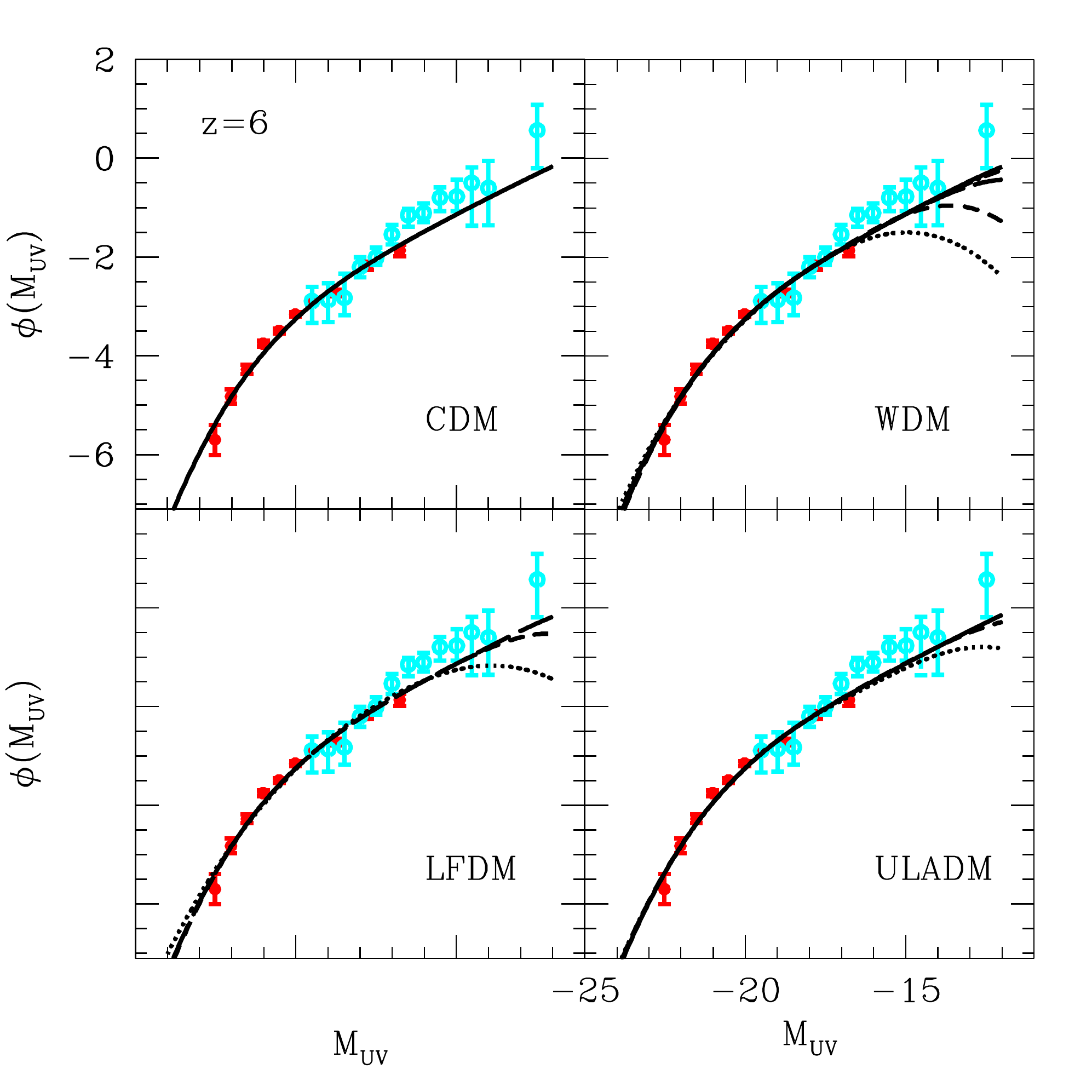}
\caption{\label{figlivz6} As in Fig.~\ref{fig9} for L16 data at $z=6$ (cyan filled circles).}
\end{figure}

Here, we discuss how the LF data at $z=6$ from L16 modify the constraints on the DM models inferred in the previous section. Since, these measurements points to a steeper faint-end slope of the LF, we may expect the constraints on the DM scenarios to favorite models with a SFR-M$_h$ relation closer to that of the CDM case. In Table~\ref{tab5} we quote the $\chi^2$ of the DM models best-fitting the LF data at $z=6$ (L16+B15 for a total 23 data points), while in Fig.~\ref{figlivz6} we plot the corresponding best-fit luminosity functions. Not surprisingly the models with the lowest $\chi^2$ values are those with the lest suppression of halo abundance at low masses. 

However, it is worth noticing that none of the models provide a good fit to the $z=6$ LF data, since the reduced $\chi^2_{\rm red}\gtrsim \mathcal{O}(2)$. This may point to the fact that the L16 data at $z=6$ requires a shallower slope of the SFR-M$_h$ at low masses than that inferred from the LF data at $z=4$ and $5$, which is not the case at $z=7$ and $8$. We find WDM-5, LFDM-2 and ULADM-3 to be the models with the lowest $\chi^2_{\rm tot}$ values with respect to the realisations of the same DM scenario.

The deviance statistics indicates that the best-fit WDM-1 model is excluded at more than $4\sigma$ with $\Delta\chi^2_{\rm tot}=90.2$, WDM-2 is excluded at more than $3\sigma$ with $\Delta\chi^2_{\rm tot}=13.3$. while WDM-3 has $\Delta\chi^2_{\rm tot}$. This suggests a lower bound on the WDM thermal relic mass corresponding to $m_{\rm WDM}\gtrsim 2.0$ keV. This is slightly stronger than that inferred using the Bouwens et al. data at $z=6$. In the case of LFDM models, the deviance statistics exclude the best-fit LFDM-1 model to more than $4\sigma$ with $\Delta\chi^2_{\rm tot}=36.0$, while LFDM-2 and LFDM-3 are within $1\sigma$ of each other. This corresponds a constraint on $z_t$ similar to the inferred in Section~\ref{result}. On the other hand for the ultra-light axion models we find that ULADM-1 and ULADM-2 are excluded at more than $2\sigma$ with $\Delta\chi^2_{\rm tot}=18.5$ and $4.9$ respectively. This points to a strong bound on the axion mass $m_a\gtrsim 1.5\times 10^{-21}$ eV at $2\sigma$, which is consistent with the constraints found in \cite{Menci2017} using the galaxy number density estimated from L16 at $z=6$.

\begin{table}[th]
\centering
\begin{tabular}{|c|c|c|}
\hline\hline
Model & $\chi^2_{z=6}$ &$\chi^2_{\rm tot}$\\
\hline
{\rm \textbf{CDM}} & 47.1 & \textbf{90.5} \\
\hline
{\rm WDM-1} & 98.2 & 179.4\\
{\rm WDM-2} & 53.4 & 102.5\\
{\rm WDM-3} & 48.6 & 92.4\\
{\rm WDM-4} & 47.0 & 90.7\\
{\rm \textbf{WDM-5}} & 46.3 & \textbf{89.2}\\
\hline
{\rm LFDM-1} & 67.8 & 129.4\\
{\rm \textbf{LFDM-2}} & 49.5 & \textbf{93.4}\\
{\rm LFDM-3} & 47.2 & 93.9\\
\hline
{\rm ULADM-1} & 59.8 & 108.3\\
{\rm ULADM-2} & 46.8 & 94.7\\
{\rm \textbf{ULADM-3}} & 46.6 & \textbf{89.8}\\
\hline\hline
\end{tabular}
\caption{\label{tab5} Best-fit value of $\chi^2_{z=6}$ and $\chi^2_{\rm tot}$ for the different DM models using L16+B15 data at $z=6$.}
\end{table}

\section{Conclusions}\label{conclude}
Measurement of the faint-end of the galaxy luminosity function at high redshifts is key to understanding the connection between early galaxy formation and scenarios of cosmic reionization. Moreover, by probing the abundance of far distant galaxies hosted in the lightest DM halos, one can also potentially test the nature of dark matter particles in the universe. Scenarios alternative to the CDM paradigm have been investigated in recent years in response to the lack of detection of weakly interacting massive particles in laboratory experiments and as a possible solution to anomalies in the observed distribution of matter at small scales.
  
Using up-to-date measurements of the high-redshift galaxy luminosity function we infer constraints on DM scenarios alternative to CDM that feature a small scale cut-off in the linear matter power spectrum. To this purpose we have run a series of high-resolution N-body simulations of warm dark matter, late-forming dark matter and ultra-light axion dark matter models to accurately characterise the low-mass end of the halo mass function at high redshifts ($z\gtrsim 4$). 

We have removed artificial groups of particles from the N-body halo catalogs using a selection criterion based on the analysis of the structural properties of the halos. We have used the resulting halo catalogs to calibrate analytical formula of the halo mass function which we have utilised to infer DM model predictions of the high-redshift galaxy luminosity function. 

In order to convert halo masses into UV-magnitudes we have developed an empirical approach based on halo abundance matching that has a twofold advantage: (i) it accounts for the effect of dust extinction which may alter the redshift dependence of the UV-magnitude halo mass relation, and (ii) it allows us to gain insight on the star formation rate of galaxies as a function of host halo mass and redshift. 

Using a compilation of state-of-art measurements of the LF at $z=6,7$ and $8$ we perform a likelihood analysis to evaluate the goodness-of-fit of the simulated DM models and infer constraints on the amplitude and scatter of the ensemble average SFR-M$_{\rm h}$ relation in such models.

We find that at fixed halo mass the average SFR slightly increases with increasing redshift, while the scatter remains constant. For all DM models considered, the SFR-M$_{\rm h}$ relation converges to the double power law behaviour of the CDM model at ${\rm M_{h}}\gtrsim 10^{12}\,{\rm M}_\odot$ h$^{-1}$, while differences occur at lower masses. In particular, DM models characterised by a suppression of low-mass halo abundance exhibit systematically higher SFR compared to the CDM scenario. This suggests that baryonic processes responsible for star formation in low-mass halos cannot be treated independently of the assumptions on the nature of the DM. Our results also indicate that independent measurements of SFR and galaxy host halo mass in this mass range and at these redshifts can directly constrain DM models.

Besides CDM, the other DM models best-fitting the LF data with lowest value of $\chi^2_{\rm tot}$ are WDM-3, LFDM-2 and ULADM-3. These are statistically indistinguishable from the best-fit CDM model, with differences $\Delta\chi^2_{\rm tot}\lesssim 1$. In contrast, we find the goodness-of-fit within the same DM scenario to vary from one model realisation to another. Thus, we infer constraints on the DM scenarios from deviance statistics. In particular, we obtain a lower bound on the WDM thermal relic particle mass $m_{\rm WDM}\gtrsim 1.5$ keV at $2\sigma$. This is less stringent than the limits found in \cite{Menci2016} which have used the L16 data at $z=6$. The LFDM models are constrained to have a phase transition redshift $z_t\gtrsim 8\cdot 10^5$ at $2\sigma$. We find ULADM best-fit models to be statistically compatible with LF data well within $2\sigma$ of the deviance statistics. 

We would like to stress that LF measurements at $z=6$ are consistent with a flattening or a turnover at faint UV-magnitudes, a point already highlighted in \cite{Bouwens2016c}. This explains as to why models such WDM-3 and LFDM-2 have $\chi^2_{\rm tot}$ values that are slightly lower than CDM. The presence of such a turnover at the faint-end of the LF has been predicted in a number of galaxy formation studies based on CDM/hydro simulations. Here, we have shown that such a feature can be a signature of non-standard DM. However, it is important to note that in such non-standard DM models as considered here, a gentle turnover requires a higher SFR at low halos masses compared to the CDM prediction to compensate for the sharp drop of halo abundance at these masses. In fact, the constraints we have derived on the particle mass in WDM and ULADM models and the phase transition redshift in LFDM would be much tighter if we had assumed as template the CDM model's average SFR-M$_{\rm h}$ relation.

The redshift evolution of the halo mass function at the low-mass end as well as the SFR histories featured by the investigated models suggest that further constraints can be inferred from the \textit{Planck} determination of optical depth \cite{PlanckReio2016} and more in general from studies of the cosmic reionization history. However, this will be possible only at the cost of additional caveats. Regarding this last point, a tomographic reconstruction of the reionization history through cross-correlation of CMB temperature and polarization maps with the angular distribution of reionization tracers as proposed in \cite{Munshi2014} can also probe DM scenarios. Similarly, for measurements of cosmic reionization history from kinetic Sunyaev-Zeldovich detections (see e.g. \cite{kSZ}). These are relevant aspects that we plan to explore in future.

\begin{acknowledgments}
P.S.C. would like to thank Rebecca Bowler, Nicola Menci, Paolo Salucci and Andrea Lapi for useful comments and discussions. D.J.E.M. acknowledges support of a Royal Astronomical Society fellowship hosted at King's College London, and useful conversations with Brandon Bozek, Jeremiah Ostriker, and Rosemary Wyse. S.A. thanks Romeel Dav\'e for discussions. This work was granted access to the HPC resources of TGCC under the allocation 2016 - 042287 made by GENCI (Grand Equipement National de Calcul Intensif) on the machine Curie. The research leading to these results has received funding from the European Research Council under the European Community Seventh Framework Programme (FP7/2007-2013 Grant Agreement no. 279954). We acknowledge support from the DIM ACAV of the Region Ile-de-France.
\end{acknowledgments}

\appendix
\section{WDM, LFDM and ULDAM halo mass function best-fit coefficients}\label{app1}

\begin{figure*}[ht]
\centering
\subfigure{\includegraphics[scale=0.25]{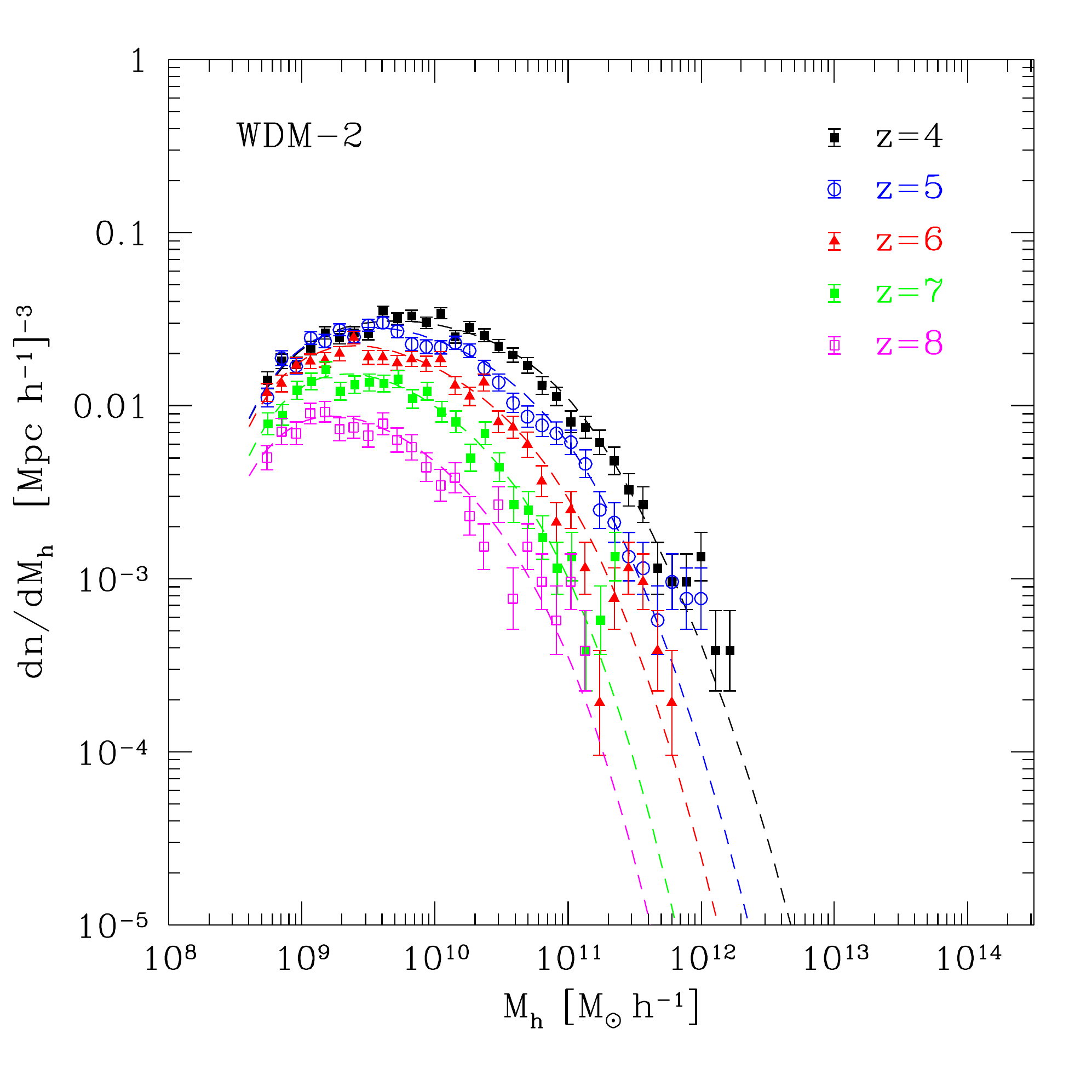}}\quad
\subfigure{\includegraphics[scale=0.25]{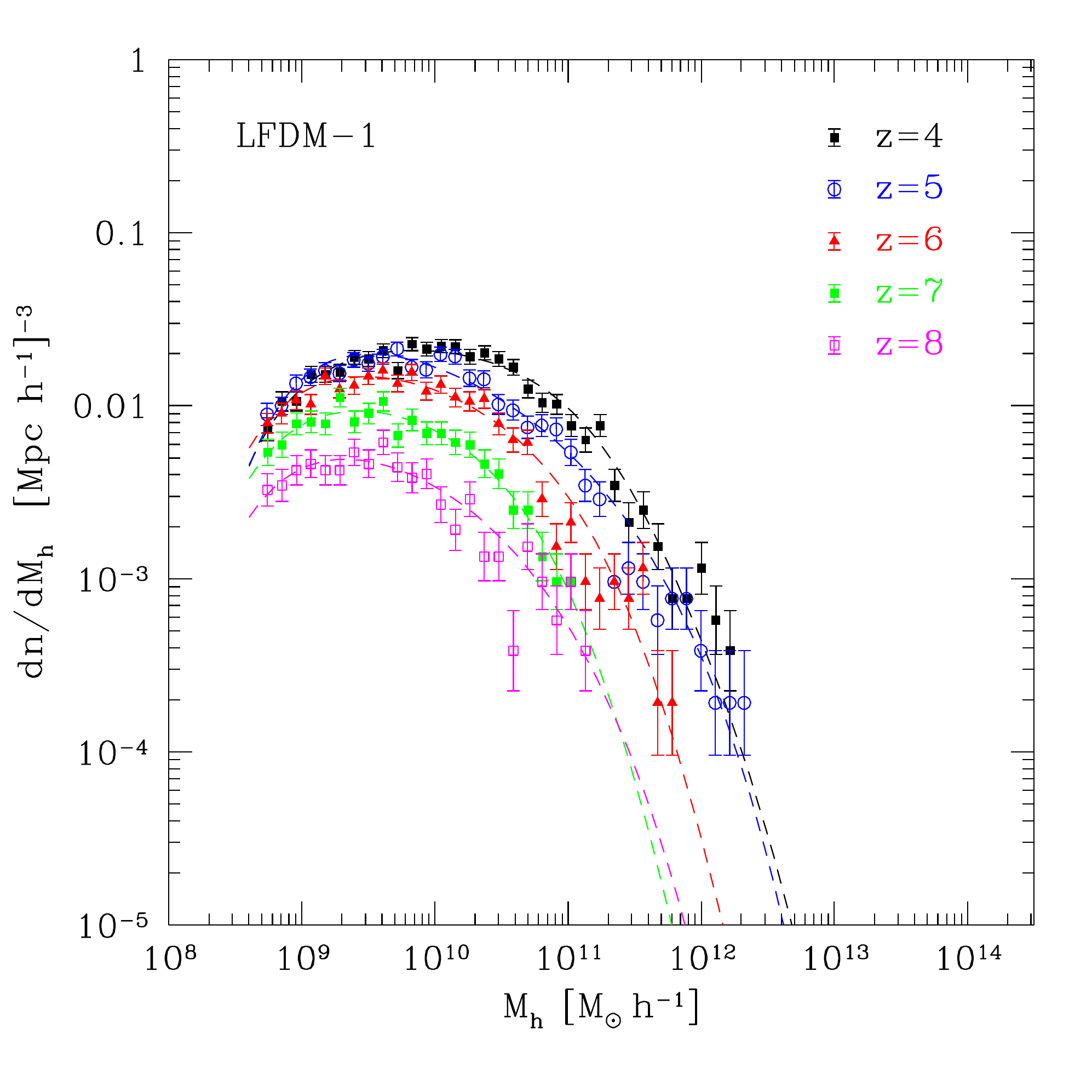}}\quad
\subfigure{\includegraphics[scale=0.25]{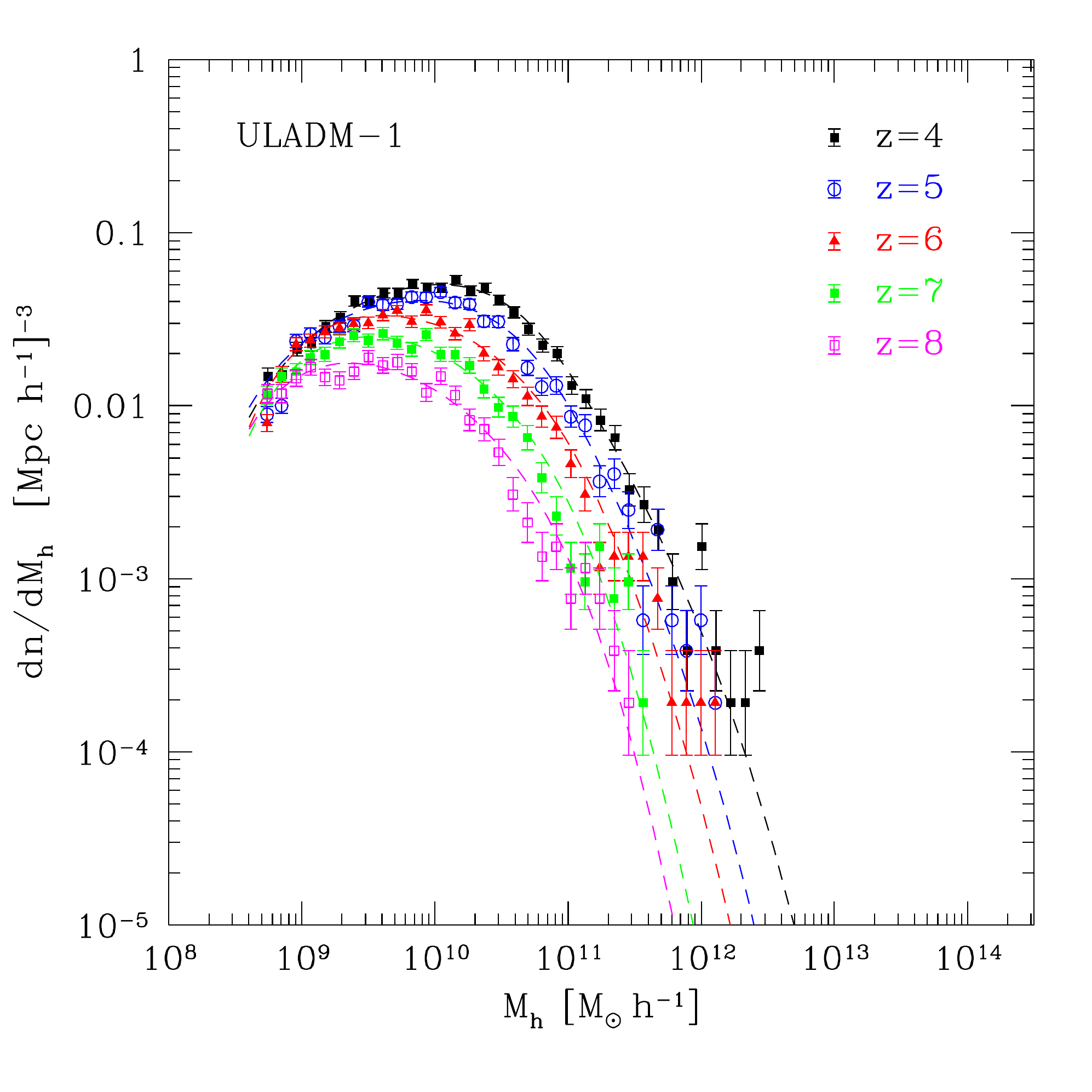}}
\caption{\label{figA1}Halo mass function from N-body halo catalogs of WDM-2 (left panel), LFDM-1 (center panel) and ULADM-1 (right panel) and best-fit functions at $z=4,5,6,7$ and $8$ (data points and curves from top to bottom)}
\end{figure*}

In Table~\ref{tabA1},~\ref{tabA2} and~\ref{tabA3} we quote the values of the coefficients of Eq.~(\ref{dndm_ndm}) best-fitting the halo mass functions from the N-body halo catalogs of WDM, LFDM and ULADM model simulations. For illustrative purposes in Fig.~\ref{figA1} we plot the mass function at $4\le z\le 8$ from the simulations of WDM-2 (panel a), LFDM-1 (panel b) and ULADM-1 (panel c) models against the best-fitting functions. 

\begin{table*}[t]
\centering
\begin{tabular}{|c|c|c|c|c|c|}
\hline\hline
Model & z & $\alpha$ & $\beta$ & $\gamma$ & ${\rm M_*}$\\
\hline
     WDM-1    & 4.0 & 0.46688E-01 & -0.15362E-02 & 0.77722E+00 & 0.36452E+12 \\
             & 5.0 & 0.56763E-01 & -0.15343E-02 & 0.77124E+00 & 0.35490E+12 \\  
             & 6.0 & 0.68210E-01 & -0.14801E-02 & 0.80043E+00 & 0.35685E+12 \\   
             & 7.0 & 0.12417E+00 & -0.10537E-02 & 0.88205E+00 & 0.33658E+12 \\    
             & 8.0 & 0.17072E+00 & -0.87254E-03 &  0.97998E+00 &  0.31569E+12 \\ 
\hline                 
     WDM-2  & 4.0 &  -0.19738E-01 &  -0.41804E-02 &  0.86551E+00 &   0.67546E+11\\
                  &  5.0 &  0.85360E-02 &  -0.39280E-02 & 0.80553E+00 &   0.79238E+11\\
                  & 6.0 &   0.85360E-02  &  -0.39280E-02 & 0.80553E+00  &  0.79238E+11\\  
                  & 7.0 &   0.11670E-01  &  -0.41489E-02 & 0.82052E+00 &   0.80275E+11 \\
                  & 8.0 & 0.11890E-01 &  -0.31869E-02 & 0.86675E+00  &  0.89083E+11\\ 
\hline
    WDM-3 & 4.0 &  -0.29145E-01 &  -0.13439E-01  & 0.53411E+00  &  0.25237E+11\\ 
                 & 5.0 &   0.21376E-02 &  -0.10081E-01  & 0.58836E+00  &  0.27828E+11 \\
                 & 6.0 &  0.10142E-01 &  -0.94886E-02 & 0.59938E+00  &  0.28294E+11 \\    
                 & 7.0 &   0.16513E-01 &  -0.97150E-02  & 0.61578E+00  &  0.28749E+11\\    
                 & 8.0 &  0.17838E-01 &  -0.97581E-02 & 0.62047E+00  &   0.28882E+11 \\    
\hline
   WDM-4  & 4.0 & -0.54437E-01 &  -0.12643E-01 &   0.51662E+00  &   0.83804E+10 \\            
                 & 5.0 & -0.19856E-01 &  -0.15134E-01  &  0.42798E+00  &  0.11073E+11\\
                 & 6.0 & 0.44430E-02  &  -0.17341E-01 &  0.53405E+00  &  0.87945E+10 \\
                 & 7.0 &  0.56581E-02  & -0.18033E-01 & 0.55238E+00  &  0.90182E+10 \\  
                 & 8.0 & 0.75246E-02 &  -0.18507E-01  & 0.56916E+00  &  0.92176E+10\\ 
\hline
       WDM-5 & 4.0 & -0.34874E-01 &  -0.14503E-01 & 0.30686E+00  &  0.88613E+10 \\  
                    & 5.0 & -0.22920E-01 &  -0.14582E-01 &  0.29235E+00  &  0.89928E+10 \\  
                    & 6.0 & -0.18762E-01 &  -0.19141E-01 &  0.38016E+00 &   0.58888E+10\\   
                    & 7.0 & 0.40597E-04 &  -0.21623E-01 & 0.41057E+00  &  0.58092E+10  \\
                    & 8.0 & 0.19086E-02 &  -0.22924E-01 & 0.43360E+00  &  0.59553E+10  \\  
\hline\hline
\end{tabular}
\caption{\label{tabA1}Halo mass function best-fit coefficients of Eq.~(\ref{dndm_ndm}) for WDM models.}
\end{table*}

\begin{table*}[t]
\centering
\begin{tabular}{|c|c|c|c|c|c|}
\hline\hline
Model & z & $\alpha$ & $\beta$ & $\gamma$ & ${\rm M_*}$\\
\hline
     LFDM-1 & 4.0 & 0.77542E-02 &  -0.30292E-02 & 0.91358E+00  &  0.98292E+11\\
                   & 5.0 &  0.14193E+01 &   -0.28145E-04 & 0.71418E+00 &   0.16774E+14 \\
                   & 6.0 &   0.12092E+00 &   -0.20581E-02 & 0.91856E+00  &  0.11639E+12 \\
                   & 7.0 &  -0.94165E-01 &  -0.38357E-02 & 0.10138E+01 &   0.57870E+11\\
                   & 8.0 & 0.28828E+01 &  -0.37645E-05 & 0.10240E+01  &  0.59584E+14\\
                   \hline
     LFDM-2 & 4.0 &   -0.29327E-01  &  -0.15755E-01 &  0.73613E+00  &  0.19674E+11 \\
                   & 5.0 &  0.97870E-03 &  -0.11420E-01 & 0.77712E+00 &   0.21978E+11\\
                   & 6.0 &  0.80251E-02  & -0.89361E-02 &  0.79105E+00  &  0.23007E+11\\
                   & 7.0 &  0.89883E-02 &  -0.90835E-02 & 0.79925E+00 &   0.23181E+11 \\
                   & 8.0 &  0.18154E-01 &  -0.95860E-02  & 0.81578E+00  &  0.23253E+11\\
                   \hline
    LFDM-3 & 4.0 &   -0.32719E-01  &  -0.20121E-01 &  0.40862E+00  &  0.68945E+10  \\
                  & 5.0 &  -0.13631E-01 &   -0.22584E-01 &  0.33976E+00  &  0.78197E+10\\
                  & 6.0 &  -0.19237E-02 &  -0.27440E-01 & 0.23097E+00 &   0.95535E+10  \\
                  & 7.0 &  0.18435E+00  & -0.57700E-02 &  0.27619E+00  &  0.43788E+11\\  
                  & 8.0 &   0.19969E+00  &  -0.82653E-02 & 0.41080E+00 &   0.23100E+11\\    
\hline\hline
\end{tabular}
\caption{\label{tabA2}Halo mass function best-fit coefficients of Eq.~(\ref{dndm_ndm}) for LFDM models.}
\end{table*}

\begin{table*}[t]
\centering
\begin{tabular}{|c|c|c|c|c|c|}
\hline\hline
Model & z & $\alpha$ & $\beta$ & $\gamma$ & ${\rm M_*}$\\
\hline
     ULADM-1 & 4.0 &  0.53670E-01  &  -0.80077E-02 & 0.11984E+01  &  0.26093E+11 \\
                     & 5.0 & 0.13686E+00 &   -0.60703E-02 & 0.11552E+01  &   0.32254E+11\\
                     & 6.0 & 0.30520E+00  & -0.53837E-02 & 0.95829E+00  &  0.61748E+11\\
                     & 7.0 & 0.46699E+00  & -0.41026E-02 &  0.10038E+01 &   0.75679E+11 \\
                     & 8.0 & 0.63537E+00  &  -0.24629E-02 &  0.10432E+01 &   0.98482E+11 \\
    
     \hline
     ULADM-2 & 4.0 & 0.16547E-01 &  -0.37448E-01 &  0.72171E+00  &  0.75500E+10\\
                     & 5.0 &   0.38302E-01  & -0.43754E-01 & 0.87168E+00 &    0.60405E+10\\
                     & 6.0 & 0.72256E-01 &  -0.27448E-01 & 0.10546E+01  &  0.59524E+10 \\
                     & 7.0 &  0.28522E+00  &  -0.13903E-01 & 0.69252E+00  &  0.20848E+11 \\
                     & 8.0 & 0.35025E+00 &  -0.11690E-01 & 0.77661E+00 &   0.21250E+11  \\        
       \hline
       ULADM-3 & 4.0 & 0.56134E-02  &  -0.93856E-01 & 0.48333E+00  &  0.12463E+10\\
                        & 5.0 & 0.15222E-01 &  -0.18837E+00 & 0.43060E+00  &  0.86619E+09\\
                        & 6.0 &  0.14859E-01 &  -0.54771E-01 & 0.62784E+00  &  0.11009E+10\\   
                        & 7.0 &  0.55854E-01 &  -0.91105E-01 & 0.36028E+00  &  0.13199E+10\\ 
                        & 8.0 &  0.84951E-01 &   -0.88151E-01 & 0.45541E+00 &   0.13210E+10\\                   
\hline\hline
\end{tabular}
\caption{\label{tabA3}Halo mass function best-fit coefficients of Eq.~(\ref{dndm_ndm}) for ULADM models.}
\end{table*}

\end{document}